\documentclass[preprint,3p,times,authoryear,12pt]{elsarticle}

\usepackage{amssymb, amsmath}
\usepackage{newtxtext, newtxmath}
\usepackage{hyperref}
\usepackage{optidef}

\usepackage{nomencl}
\usepackage{float, array, colortbl}
\usepackage{multirow, multicol}
\usepackage{algorithm}
\usepackage[noend]{algpseudocode}
\usepackage{siunitx, booktabs}
\usepackage{empheq}
\usepackage[dvipsnames]{xcolor}
\usepackage{makecell}

\hypersetup{
    colorlinks,
    linkcolor={red!50!black},
    citecolor={blue!50!black},
    urlcolor={blue!80!black}
}
\usepackage{bm,fix-cm}
\DeclareMathAlphabet{\mathsfit}{T1}{\sfdefault}{\mddefault}{\updefault}
\SetMathAlphabet{\mathsfit}{bold}{T1}{\sfdefault}{\bfdefault}{\updefault}

\newcommand{\vect}[1]{{\bm{\mathit{#1}}}}
\newcommand{\matr}[1]{\bm{\mathsfit{#1}}}
\newcommand{\tensor}[1]{{\mathsfit{#1}}}

\newcommand{\mtxt}[1]{\ensuremath{\mathrm{#1}}}
\newcommand{\totald}{\ensuremath{\mathrm{d}}\,}
\newcommand{\T}{^\mathrm{T}}
\newcommand{\E}[1]{\ensuremath{\cdot 10^{#1}}}
\newcommand{\overbar}[1]{\mkern1.5mu\overline{\mkern-2mu#1\mkern-1mu}\mkern 1.3mu}

\newlength{\nomitemorigsep}
\makenomenclature

\makeatletter
\newif\if@nomlist

\newcommand*\nomlist{%
  \@nomlisttrue
  \list{}{%
    \labelwidth\nom@tempdim
    \leftmargin\labelwidth
    \advance\leftmargin\labelsep
    \itemsep\nomitemsep
    \let\makelabel\nomlabel}}

\renewcommand*\thenomenclature{%
  \@ifundefined{chapter}%
    {\section*{\nomname}\if@intoc\addcontentsline{toc}{section}{\nomname}\fi}%
    {\chapter*{\nomname}\if@intoc\addcontentsline{toc}{chapter}{\nomname}\fi}%
  \nompreamble
  \@nomlistfalse
}

\renewcommand\nomgroup[1]{%
  \if@nomlist\endlist\end{multicols}\fi
  \ifx#1A\relax
    \def\nomgroupname{Data assimilation symbols}%
  \else
    \ifx#1B\relax
      \def\nomgroupname{Echo state network symbols}%
    \else
      \def\nomgroupname{Superscripts/Subscripts}%
    \fi
  \fi
  \begin{multicols}{2}[\raggedcolumns\noindent\textit{\nomgroupname}]
  \nomlist
}

\renewcommand*\nomname{}
\renewcommand*\nompreamble{\vspace*{-6\parsep}}
\renewcommand*\nompostamble{\end{multicols}}
\makeatother

\newcommand{\toLM}[1]{#1}% AN to LM
\newcommand{\revone}[1]{#1}% Comments for rev1
\newcommand{\revtwo}[1]{#1}% Comments for rev2

\begin{document}

\begin{frontmatter}

\title{Inferring unknown unknowns: \\ Regularized bias-aware ensemble Kalman filter}

\author[inst1,inst3]{Andrea Nóvoa}
\author[inst2,inst3,inst1]{Alberto Racca}
\author[inst3,inst4,inst1]{Luca Magri}

\affiliation[inst1]{organization={Cambridge University Engineering Dept.},
            addressline={Trumpington St}, 
            city={Cambridge},
            postcode={CB2 1PZ},
            country={UK}}

\affiliation[inst2]{organization={Imperial College London, I-X Centre for AI In Science},%Department and Organization
            addressline={White City Campus, 84 Wood Lane}, 
            city={London},
            postcode={W12 7SL}, 
            country={UK}}

\affiliation[inst3]{organization={Imperial College London, Aeronautics Dept.},%Department and Organization
            addressline={Exhibition Road}, 
            city={London},
            postcode={SW7 2AZ}, 
            country={UK}}
\affiliation[inst4]{organization={The Alan Turing Institute},
            addressline={96 Euston Rd}, 
            city={London},
            postcode={NW1 2DB}, 
            country={UK}}

\begin{abstract}
%The terms `bias', `systematic error', and `model error' are equivalent to each other for the purpose of this paper, i.e.,~they are informally referred to as the `unknown unknowns'.
Because of physical assumptions and numerical approximations, low-order models are affected by uncertainties in the state and parameters, and by model biases.
% 
%On the one hand, data assimilation enables the estimation of the low-order model's states and parameters by combining model predictions with data from sensors.  
%% 
%On the other hand, 
Model biases, also known as model errors or systematic errors, are difficult to infer because they are `unknown unknowns', i.e.,~we do not necessarily know their functional form \textit{a priori}. 
With biased models, data assimilation methods may be ill-posed because either 
(i) they are `bias-unaware' because the estimators are assumed unbiased, 
(ii) they rely on an \textit{a priori} parametric model for the bias, or 
(iii) they can infer model biases that are not unique for the same model and data. 
% 
%Whereas the inference of model parameters and state is well established in data assimilation, 
%Real-time modelling often employs low-order models, which may contain model biases, thus the analysis from traditional data assimilation are sub-optimal.  
%The estimation of model biases
% 
% 
%Current bias-aware frameworks face two limitations: 
%(i) they rely on an \textit{a priori} definition of a parametric model of the bias whose parameters are inferred within the assimilation; and 
% (ii) they cannot ensure that the norm of the bias is reduced in the analysis \ar{do you ensure it here? is it not a soft constraint? maybe ’nudge’ or similar}
%\lm{(ii) they indistinguishably recover analysis states with a reduced or increased bias.} 
%
%The overarching objective of this work is to overcome these limitations.  
% 
First, we design a data assimilation framework to perform combined state,  parameter, and  bias estimation. 
Second, we propose a mathematical solution with a sequential method, i.e.,~the \textit{regularized bias-aware ensemble Kalman Filter} (r-EnKF), which requires a model of the bias and its gradient (i.e., the Jacobian). 
Third, we propose an echo state network as the model bias estimator. We  derive the Jacobian of the network, and design a robust training strategy with data augmentation to accurately infer the bias in different scenarios. 
Fourth, we apply the r-EnKF to nonlinearly coupled oscillators (with and without time-delay) affected by different forms of bias. The r-EnKF infers in real-time parameters and states, and a unique bias.
The applications that we showcase are relevant to acoustics, thermoacoustics, and vibrations; however, the r-EnKF  opens new opportunities for combined state, parameter and bias estimation for \revone{real-time and on-the-fly prediction in nonlinear systems. }

\end{abstract}

% %%Graphical abstract
% \begin{graphicalabstract}
% \end{graphicalabstract}

% %%Research highlights
% \begin{highlights}
% \item Research highlight 1
% \item Research highlight 2
% \end{highlights}


\end{frontmatter}

%% \linenumbers

\section{Introduction}
\label{sec:intro}

% Data Assimilation and bias 
Data Assimilation (DA) optimally combines uncertain measurements  with \revone{an imperfect numerical model of the physical quantity of interest} in order to find the best estimate of the full physical state and/or  model parameters~\citep[e.g.,][]{tarantola2005inverse, evensen_data_2009}.   
DA has been developed in the last decades in the earth sciences and \revtwo{numerical weather prediction (NWP)} communities by combining concepts from control theory, dynamic programming, and Bayesian statistics.
In the NWP terminology, DA obtains the most likely estimates of the state and/or model parameters (the \textit{analysis}) that are compatible with the background knowledge from the numerical model (the \textit{forecast}), and a set of measured data (the \textit{observations}), as well as with their corresponding uncertainties. 
%

% Ṣystematic errors and bias-unaware methods
The assimilation framework can be affected by  errors in the observations (e.g., due to poorly calibrated or malfunctioning sensors), and/or forecast model errors (e.g., model assumptions, or discretization errors). If the errors are non-aleatoric or with non-zero mean, they are systematic errors, which are also known as \textit{bias}, and they may be a function of the state, space, time, and/or the environment. 
\revone{The terms `bias', `systematic error', and `model error' are equivalent to each other for the purpose of this paper, i.e.,~they are informally referred to as the `unknown unknowns'. } 
Biases in the assimilation can be detected by monitoring the long-term statistics of the \textit{innovations}, \revtwo{which are the residuals between the observations and the forecast equivalents}~\citep{haimberger_homogenization_2007}. If the innovations are not Gaussian in time, the assimilation framework may become biased, and the analysis resulting from a DA scheme may be sub-optimal~\citep{anderson_optimal_2012}. 
Therefore, the bias must be explicitly and carefully handled in operational DA. 
Nonetheless, traditional DA is designed to adjust the estimates with respect to unbiased Gaussian errors in the model and observations, thus, we refer to these as \textit{bias-unaware} methods. 
If unbiased observations are assimilated into a biased model with a bias-unaware method, the resulting analysis will be sub-optimal~\citep[e.g.,][]{dee_bias_2005, novoa_magri_2022}. Moreover, a biased model  can lead to increased analysis errors, which may result in an analysis varying with the assimilation frequency~\citep{tsyrulnikov_stochastic_2005}.  \\

Current \textit{bias-aware} DA schemes include bias estimation methods that are either  
(i) off-line, which perform first, bias-unaware DA to estimate the bias from the forecast innovations, and second, a bias-aware DA by correcting the forecast with the bias estimate from the bias-unaware DA; 
or (ii) adaptive, which estimate the bias as part of the assimilation cycle to remove the bias of the analysis.  
Adaptive bias-aware schemes are more versatile because they are robust to transient changes in the bias form~\citep{eyre_observation_2016}. 
In this work, we focus on sequential adaptive schemes.  
\revtwo{
Bias-aware methods have been mainly implemented in variational DA. 
For instance, in NWP the four-dimensional variational (4D-Var) method was extended to consider explicitly an adaptive model bias in the variational problem.
The bias correction in 4D-Var has been deployed in the model space~\citep[e.g.][]{tremolet_accounting_2006, laloyaux_towards_2020}, as well as in the observation space~\citep[e.g.][]{derber_use_1998, auligne_adaptive_2007}. Bias correction in the observations  is currently operational at the European Centre for Medium-range Weather Forecasting~\citep{han_constrained_2016}. 
}
Other applications of variational adaptive bias correction schemes can be found in ocean studies~\citep[e.g.][]{bell_assimilation_2004, balmaseda_multivariate_2007}. 
\\

%  Sequential DA and BADA
Sequential assimilation, also known as real-time DA~\citep[e.g.,][]{novoa_magri_2022}, bypasses the need of storing and post-processing  batches of data by assimilating the observations on the fly, which makes them suitable for real-time prediction and control applications. 
Real-time state and/or parameter estimation is often implemented in a Kalman filter framework~\citep{kalman_new_1960}, which provides the optimal solution in a Bayesian framework for a linear system. 
Several modifications to the filter allow us to extend and generalize the Kalman filter to nonlinear systems and  nonlinear state spaces, such as the extended Kalman filter, particle filters, or ensemble filtering methods~\citep[e.g.,][]{evensen_data_2009}.
Nevertheless, these  are \textit{bias-unaware} formulations.
Bias estimation in real-time schemes dates back to \citet{friedland_treatment_1969}, who proposed a \textit{separate-bias} Kalman filter scheme, which consists of solving two Kalman filter problems, one for the physical model state and parameter estimation, and another for estimating the parameters of the bias model. The separate-bias Kalman filter, which was initially limited to an adaptive linear biases, was generalized to handle  nonlinear and stochastic bias models~\citep[e.g.][]{friedland_notes_1978, ignagni_separate_1990, zhou_extension_1993}. Further, \citet{dee_data_1998} extended the separate filter by coupling the two sets of problems using the updated bias estimates to remove the bias of the forecast during the assimilation. 
However, the coupled filter can become costly for multivariate  models because it requires two analysis steps~\citep{balmaseda_multivariate_2007}.
\revone{
\citet{rubio2019real} proposed a sequential DA with bias correction in a Bayesian framework, which combined reduced-order models and transport map sampling. 
}
More recently, \citet{dasilva_flow_2020} deployed an ensemble Kalman filter for aerodynamic flows in which the system dynamics are augmented with a low-rank representation of the flow discretization error. By inferring the parameters of the low-rank model, they could perform a bias-aware estimation of the flow-field under significant discretization bias (which is a subset of the model bias that we are set to tackle in this paper).
The need for a pre-defined model for the bias means that the accuracy of the methods is bounded to the complexity and suitability of this pre-defined model. Further, the increase in the number of unknown parameters to infer can affect the conditioning of the minimization problem~\citep{dee_bias_2005}.
\\

\revone{
Alternatively, hybrid DA methods, which decouple the estimation of the model bias from the  assimilation framework to allow more flexibility and complexity in the model of the bias have been deployed in computational mechanics~\citep[e.g.,][]{chinesta_virtual_2020}. 
These hybrid methods estimate the model bias using data-driven model identification tools and Kalman filter frameworks~\citep[e.g.][]{marchand_real_2016, astroza_dual_2019, diaz_new_2023}. \citet{haik_real_2023} proposed a parameterized-background-data-weak technique to estimate the bias introduced with model-order reduction tools.
}
Recent advances in machine learning for data-driven modelling allow us to develop surrogate models of dynamical systems using neural networks, either as the full representation of the dynamics,  or on top of the pre-existing imperfect physical model as a hybrid model~\citep[e.g.,][]{abarbanel_machine_2018,watson_applying_2019, chinesta_virtual_2020}. The hybrid approach is particularly attractive from a model-bias inference perspective because neural networks can be deployed to close the low-order model, i.e.,~to estimate the bias of the model. 
%
% Combining ML & DA
The similarities between machine learning and data assimilation have recently encouraged the combination of both inverse problems~\citep{bocquet_bayesian_2020, geer_learning_2021}.
For instance, \citet{brajard_combining_2020} applied an ensemble Kalman filter to a surrogate model provided by a convolutional neural network, to allow the latter to work under sparse observations.  
In model bias applications, \citet{bonavita_machine_2020} implemented a fully-connected artificial neural network for atmospheric model bias correction in a weakly-constraint 4D-Var framework; and \citet{farchi_using_2021} proposed a network with a combination of convolutional and fully-connected layers in the same variational framework.
Training convolutional and fully connected networks may be computationally expensive because they use back-propagation, which makes them expensive for sequential data assimilation. %
In real-time assimilation, \citet{novoa_magri_2022} explained that echo state networks, which are a subset of reservoir computers, are more suitable for real-time DA  because their training consists of a computationally-cheap linear regression problem.
 % 
 % They are generalized nonlinear auto-regressive functions~\citep[][p. 306]{aggarwal2018neural}.   
%
\citet{novoa_magri_2022} employed an echo state network to provide an estimate of the bias of a low-order model, and correct the forecast state prior to applying a square-root ensemble Kalman filter~\citep{evensen_data_2009}. 
They showed that the trained network can estimate in time the bias of the state without the need to infer additional parameters. 
However, even if we have a universal approximator acting as a closure model such that observation data is assimilated into a bias-corrected state, current bias-aware schemes do not ensure that the bias is unique, \revtwo{and the analysis could lead to unrealistically-large bias estimates~\citep{han_constrained_2016}. }
Indeed, current filters do not have information on the magnitude of the bias of the forecast (its norm), thus, there is no constraint on the evolution of the bias. This means that the model state and parameters may be markedly inaccurate if the bias in not uniquely inferred. \\

% Limitations of current BADA and ML/DA approaches
\toLM{
To summarize, the available adaptive bias-aware methods face two limitations when dealing with unknown biases: (1) they rely on an \textit{a priori} definition of a parameterized model for the bias whose parameters are inferred within the assimilation framework, and (2) they do not provide a unique estimate of the bias.  
% 
% Objectives 
The overarching objective of this work is to generalize current bias-aware data assimilation methods to perform real-time state, parameter and model bias estimation of low-order models. 
We implement a three-fold strategy: 
first, we propose a regularized bias-aware sequential DA framework;
second, we provide mathematical solution to the framework with a sequential method, which performs on-the-fly state, parameter and model bias inference, whilst providing a solution with a unique model bias; and 
third, we generalize the echo state network implementation of \citet{novoa_magri_2022} for a more versatile and robust bias estimator. 
We test the proposed method in nonlinearly coupled oscillators with and without time delays, in which we consider different biases.\\
}

% Structure 
This paper is divided into seven sections, including this introduction.  
Section~\ref{sec:SDA} formalizes the problem and provides an overview of bias-unaware and bias-aware sequential DA.  
Section~\ref{sec:r-EnKF} introduces and derives the proposed regularized bias-aware ensemble Kalman filter.
Section~\ref{sec:ESN} describes the echo state network deployed for model bias estimation and the training approach; \revone{and details the derivation of the echo state network's Jacobian.} 
Section~\ref{sec:implementation} describes the implementation of the developed method  for combining the proposed data assimilation framework with an echo state network. 
In Section~\ref{sec:test_cases}, we test the proposed framework in two numerical scenarios.  
Section~\ref{sec:conclusions} ends the paper with the conclusions.

%======================================================================================================================================================================================================================================================================================================================== 
\section{Sequential data assimilation for low-order models}\label{sec:SDA}
This section  provides the motivation as well as the background knowledge on data assimilation required for the proposed regularized bias-aware sequential data assimilation framework for low-order models, which is derived in~\S~\ref{sec:r-EnKF}.
We define the model in~\S~\ref{sec:problem_statement}, and cast the framework to an ensemble framework into~\S~\ref{sec:ensemble_framework}. 
We provide an overview of the bias-unaware ensemble Kalman filter in~\S~\ref{sec:BB} and the current bias-aware assimilation practice in~\S~\ref{sec:BAEnKF}. 

\subsection{The model}\label{sec:problem_statement}
We consider a deterministic dynamical system 
\begin{align}\nonumber
\dfrac{\totald \vect{\phi}}{\totald t}&=\mathcal{F}\left(\vect{\phi}, \vect{\alpha} \right),  \\ \label{eq:model0}
\vect{y}&=\mathcal{M}(\vect{x}, \vect{\phi})+\vect{b}, 
\end{align}
where
$\vect{y}\in\mathbb{R}^{N_q}$ are the model \textit{observables}, which describe the dynamics of a measurable physical quantity;  
$\mathcal{F}:\mathbb{R}^{N_\phi}\rightarrow\mathbb{R}^{N_\phi}$ is a nonlinear operator, which is a function of the model parameters $\vect{\alpha}\in\mathbb{R}^{N_\alpha}$ and operates on the  state variables $\vect{\phi}\in\mathbb{R}^{N_\phi}$;  
$\mathcal{M}:\mathbb{R}^{N_\phi}\rightarrow\mathbb{R}^{N_q}$ is the measurement operator that maps the state variables into the observable space at the spatial locations~$\vect{x}$;  
and $\vect{b}\in\mathbb{R}^{N_q}$ is the bias. 
(The terms `bias', `systematic error', and `model error' are equivalent to each other for the purpose of this paper, i.e.,~they are informally referred to as the `unknown unknowns'. In this paper, we will use the term `bias'.) 
The bias may arise from modelling assumptions, and from approximations in the operators $\mathcal{F}$ and $\mathcal{M}$.  We define the bias of the observable space because this is the space in which the error is measurable. 
$\mathcal{F}$ can represent the spatial discretization of a partial differential equation and its boundary conditions. 

\revtwo{In reality, numerical models are subject to uncertainties in the state and parameters, which are commonly modelled as stochastic processes. }
\toLM{ 
Data assimilation can be used to improve our knowledge of the model state and parameters. This process is known as combined state and parameter estimation, which typically requires reformulating the problem with an augmented state vector comprising of the model state variables and the poorly-known parameters~\citep[e.g.,][]{yu_combined_2019}. 
For reasons that will become apparent in~\S~\ref{sec:ensemble_framework}, we augment further the state vector with the mapped model observables, such that, we define an augmented state vector $\vect{\psi}=[\vect{\phi}; \vect{\alpha}; \mathcal{M}(\vect{x}, \vect{\phi})]\in\mathbb{R}^{N=N_\phi+N_\alpha+N_q}$, where the operator $[~;~]$ indicates horizontal concatenation. }
This allows us to formally have a linear measurement operator $\matr{M}$ in the problem definition, such that \eqref{eq:model0} can be written as

\revtwo{
\begin{align}\nonumber
% \an{[\text{remove}]\dfrac{\totald}{\totald t}\begin{bmatrix}
%      \vect{\phi}\\ \vect{\alpha} \\ \mathcal{M}(\vect{x}, \vect{\phi})
%  \end{bmatrix} = \begin{bmatrix}
%       \mathcal{F} \\ \vect{0} \\ \vect{0}
%  \end{bmatrix} + \begin{bmatrix}
%      \vect{\epsilon}_\phi \\ \vect{\epsilon}_\alpha \\ \vect{\epsilon}_y
%  \end{bmatrix}}
 % \quad\therefore \quad 
 \dfrac{\totald\vect{\psi}}{\totald t} &= \matr{F}\left(\vect{\psi} \right) +\vect{\epsilon}_\psi,  \\ \label{eq:problem}
\vect{y} &= \matr{M}\vect{\psi} + \vect{b}  + \vect{\epsilon}_{y}  
\end{align}
where 
$\matr{F}(\vect{\psi}) = [\mathcal{F}(\vect{\phi},\vect{\alpha}); \vect{0}]$ is the nonlinear operator augmented with the vector of zeroes $\vect{0}\in\mathbb{R}^{N_\alpha+N_q}$ (as the parameters and observation operator are constant in time)}; 
$\vect{\epsilon}_\psi=[\vect{\epsilon}_\phi; \vect{\epsilon}_\alpha; \vect{\epsilon}_y]$ is Gaussian noise that models the aleatoric uncertainties in the augmented state, which  encapsulates the uncertainties  
in the state variables, $\vect{\epsilon}_\phi$, 
in the model parameters, $\vect{\epsilon}_\alpha$, 
and in the measurement operator, $\vect{\epsilon}_y$;
and 
$\matr{M} = \left[\matr{0}~\big|~ \mathbb{I}_{N_q}\right]$ is the vertical concatenation of a matrix of zeros, $\matr{0}\in\mathbb{R}^{N_q\times (N_\phi+N_\alpha)}$, and the identity matrix, $\mathbb{I}_{N_q}\in\mathbb{R}^{N_q\times N_q}$. 
In this work, we refer to $\matr{M}\vect{\psi}$ as the `biased' model prediction, and to $\vect{y}$ as the `unbiased' model prediction.
\newline

Sequential DA updates the model state $\vect{\phi}$ and the parameters $\vect{\alpha}$ when observation data of the modelled physical quantity are available (e.g. from experiments) by combining the observations with the model estimate $\vect{y}$. 
In a sequential framework, the observation data at time $t$, $\vect{d}(t)$, is a single measurement realization of the quantity of interest subject to uncertainty, $\vect{\epsilon}_d$, 
\begin{equation}\label{eq:epsilon_d}
    \vect{d}(t) = \vect{d}^\mtxt{t}(t) + \vect{\epsilon}_d,
\end{equation}
where the superscript `$\mtxt{t}$' stands for `true', i.e.,~$\vect{d}^\mtxt{t}(t)$ is the true physical quantity (which we do not know).   
We assume that the observations are unbiased and Gaussian distributed, i.e. $\vect{\epsilon}_d\sim\mathcal{N}\left(\vect{0},\matr{C}_{dd}\right)$, where $\mathcal{N}$ denotes the normal distribution with zero mean and covariance matrix $\matr{C}_{dd}$, \toLM{which is pre-defined as a diagonal matrix assuming that the observations are uncorrelated}.
Thus, the expectation of the observables coincides with the true physical state. 
In a real scenario, this would mean that the experimental devices are calibrated correctly, which is the overarching working assumption in this paper.  
In this numerical study, we use synthetic observable data generated by adding noise to the \textit{truth} (see~\S~\ref{sec:VdP}-\ref{sec:Rijke}). 
Lastly, we define the bias of the observable space as the difference between the truth and the expectation of the model, i.e.,~
\begin{equation}\label{eq:expected_bias}
   \vect{b} = \mathbb{E}(\vect{y}) - \mathbb{E}(\matr{M}\vect{\psi})=\vect{d}^\mtxt{t} - \mathbb{E}(\matr{M}\vect{\psi}). 
\end{equation}

% ======================================================================================== 
\subsection{Ensemble data assimilation framework}\label{sec:ensemble_framework}

We tackle the DA problem from a Bayesian framework. Our knowledge in a Markovian  model prediction  at time $t_k$ is quantified by the conditional probability  $\vect{\psi}_k \sim \mathcal{P}(\vect{\psi}_k | \vect{\psi}_{k-1}, \matr{F})$.
This knowledge is updated every time at which we have reference data from observations $\vect{d}_k$ by using the Bayes' rule, such that the \textit{posterior}  (i.e., the confidence in our prediction $\vect{\psi}_k$ after observing $\vect{d}_k$) is given  by
\begin{align}\label{eq:bayup}
\mathcal{P}(\vect{\psi}_k | \vect{d}_k,\vect{b}_k, \matr{F}) &\propto \mathcal{P}(\vect{d}_k | \vect{\psi}_k, \vect{b}_k, \matr{F}) \mathcal{P}(\vect{\psi}_k,  \vect{b}_k, \matr{F}), 
\end{align}
where 
$\mathcal{P}(\vect{\psi}_k, \vect{b}_k, \matr{F})$ is the \textit{prior}, which measures the confidence in our model's estimate if there were no available observables $\vect{d}_k$; and 
 $\mathcal{P}(\vect{d}_k | \vect{\psi}_k, \vect{b}_k, \matr{F})$ is the \textit{likelihood}, which measures the confidence that we have in the model prediction. 
The mode, i.e.,~most probable value, of $\vect{\psi}_k$ in the posterior is defined as the best estimator of the state, which is  the \textit{analysis} state. 
In other words, the analysis is the statistically consistent combination of prior physical knowledge and observed data, accounting for the model mismatch.

If the operator $\matr{F}$ were linear, the mode of the Bayesian update~\eqref{eq:bayup} would be exactly corrected by the Kalman filter equations~\citep{kalman_new_1960}. For nonlinear cases, a  common strategy consists of reformulating the problem with an ensemble approach, in which the prior is  estimated by evolving in time a number of $m$ realizations of the state. The ensemble framework is widely implemented because it avoids the computation of adjoint solvers  (which are needed in variational DA) and does not need to propagate the covariance matrix in time in contrast to, e.g., the extended Kalman filter~\citep[e.g.,][]{gonzalez_model_2017}.  
The expected value and the covariance matrix of the state $\matr{C}_{\psi\psi}$ are estimated by the ensemble statistics
\begin{equation}\label{eq:ensemble_stat}
	 \mathbb{E}(\vect{\psi})\approx\overbar{\vect{\psi}}=\dfrac{1}{m}\sum^m_{j=1}{\vect{\psi}_j} 
	 \quad \mtxt{and} \quad
	 \matr{C}_{\psi\psi} = \begin{bmatrix}
				\matr{C}_{\phi\phi}  & \matr{C}_{\phi\alpha}& \matr{C}_{\phi q} \\
				\matr{C}_{\alpha \phi}  & \matr{C}_{\alpha\alpha}& \matr{C}_{\alpha q} \\
				\matr{C}_{q \phi}  & \matr{C}_{q \alpha}& \matr{C}_{q q} \\
			\end{bmatrix}
			\approx\dfrac{1}{m-1}\sum^m_{j=1}(\vect{\psi}_i-\overbar{\vect{\psi}})\otimes(\vect{\psi}_i-\overbar{\vect{\psi}}),
\end{equation}
where $\otimes$ is the dyadic product. 
% 
%The sampling probability distribution that maximizes the information
%entropy for the first and second moments is the Gaussian distribution~\citep{jaynes_information_1957}. Therefore, 
We generate the initial ensemble from a Gaussian distribution, i.e,  $\vect{\psi}_j\sim\mathcal{N}(\overbar{\vect{\psi}}, \matr{C}_{\psi\psi})$. 
Each member in the ensemble is forecast individually as
\begin{align}\nonumber
  \dfrac{\totald \vect{\psi}_j}{\totald t} &= \matr{F}(\vect{\psi}_j)   + \vect{\epsilon}_\psi, 
\\ \label{eq:ModeForecasrEnsemble}
\vect{y}_j &= \matr{M}\vect{\psi}_j + \vect{b}_j  + \vect{\epsilon}_y , \qquad \mtxt{for} \quad j=0,\dots,m-1, 
\end{align}
where $\vect{b}_j$ is the bias of each ensemble member. 
Because the forecast operator $\mathcal{F}$ is nonlinear, the Gaussian prior may not remain Gaussian after the model forecast, and $\mathbb{E}(\mathcal{F}(\vect{\psi}))\neq\mathcal{F}(\overbar{\vect{\psi}})$. However, we work under the assumption that the time between analyses $\Delta t_\mtxt{d}$ is small enough such that the Gaussian distribution is not significantly distorted, as assumed in nonlinear DA problems~\citep{evensen_data_2009}. Therefore, the ensemble filter is a nonlinear model that approximates the mode of the Bayesian update.  
% 

%========================================================================================
\subsection{Bias-unaware Ensemble Kalman Filter}\label{sec:BB}
For each member in the ensemble, the ensemble Kalman filter (EnKF) minimizes the uncertainty in the augmented state (by reducing the covariance) and the Mahalanobis distance between the model forecast and the measurements. The cost function for each ensemble member is~\citep[e.g.,][]{reich_probabilistic_2015}
\begin{align}\label{eq:BB_cost_func}
\mathcal{J}(\vect{\psi}_j) = &\left\|\vect{\psi}_j-\vect{\psi}_j^\mtxt{f}\right\|^2_{\matr{C}^{\mtxt{f}^{-1}}_{\psi\psi}} +
 \left\|\matr{M}\vect{\psi}_j-\vect{d}\right\|^2_{\matr{C}^{-1}_{dd}}, \quad j=0,\dots,m-1.
\end{align}
where the superscript `$\mtxt{f}$' stands for `forecast', and the precision matrices $\matr{C}^{-1}$ weigh the $L_2$-norms. The state covariance matrix, $\matr{C}_{\psi\psi}$, is computed via Eq. \eqref{eq:ensemble_stat}. 
The observation error covariance, $\matr{C}_{dd}$, is pre-defined from the knowledge of the source of the data (e.g. precision  of the measuring devices).  
Each ensemble member must be updated with a different observations to avoid the underestimation of the analysis covariance~\citep{burgers_analysis_1998}. This can be achieved by adding Gaussian noise to the observables such that each $j$ ensemble member is assimilated with 
$\vect{d}_j \sim \mathcal{N}\left(\vect{d},\matr{C}_{dd}\right)$, where $\matr{C}_{dd}=\overbar{\vect{\epsilon}_d\otimes\vect{\epsilon}_d}$.

An analysis state $\vect{\psi}_j^\mtxt{a}$ minimizes \eqref{eq:BB_cost_func} if the derivative of the cost function with respect to the state is zero, which leads to the  EnKF equations~\citep{evensen_ensemble_2003}
\begin{align}\label{eq:EnKF}
    \vect{\psi}_j^\mtxt{a} 
    &= \vect{\psi}_j^\mtxt{f}+\matr{C}_{\psi\psi}^\mtxt{f}\matr{M}\T\left(\matr{C}_{dd}+\matr{M}\matr{C}_{\psi\psi}^\mtxt{f}\matr{M}\T\right)^{-1}\left[\vect{d}_j - \matr{M}\vect{\psi}_j^\mtxt{f}\right], \quad j=0,\dots,m-1,
\end{align}
where the superscript `$\mtxt{a}$' stands for `analysis'. 
From~\eqref{eq:EnKF} it can be seen that the updates to the forecast are confined to the range of the  forecast ensemble error covariance matrix    $\matr{C}_{\psi\psi}^\mtxt{f}$. 
Therefore, the size and spread of the ensemble are  key to the performance of the filter. 
In the limit of an infinite ensemble, the EnKF~\eqref{eq:EnKF} is no longer an approximation, and if the operator $\matr{F}$ is unbiased,
the mean of the analysis ensemble would recover the true state and parameters~\citep{evensen_data_2009}.
However, in practice there is a trade-off between the bias of a model (its accuracy) and its computational requirements. In fact, sequential DA methods are typically employed in real-time and on-the-fly prediction applications, which often use computationally-cheap, thus, potentially biased models~\citep[e.g.][]{yu_data_2019}. 
%

%========================================================================================
\subsection{Bias-aware ensemble Kalman filter}\label{sec:BAEnKF}
\begin{figure}[!htb]
    \centering
    \includegraphics[width=.9\textwidth]{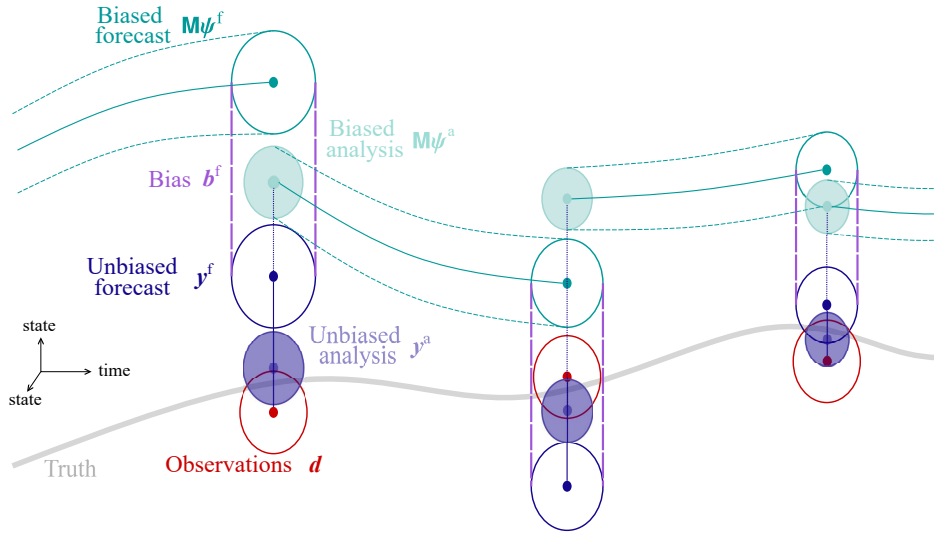}
    \caption{Conceptual schematic of a sequential bias-aware data assimilation process.
    Bias-aware DA aims to estimate the truth (grey line) by combining measurable  observations (red) and their uncertainties (represented by the circles) with the predictions from a biased model, which propagates in time (teal lines). 
    To assimilate the observations into the biased model, we must provide an estimate of the bias of the forecast (vertical violet dashed lines), which maps the biased forecast and its uncertainty (teal) into an unbiased state (navy).   
    The assimilation between the observations and the unbiased forecast results in the unbiased analysis with a reduced uncertainty (shaded navy). This analysis indirectly updates the biased forecast to a biased analysis (shaded teal), which serves as the updated initial condition for the next forecast step. 
    This  process is repeated sequentially in time when observations are available.}
    \label{fig:tubes_BADA}
\end{figure}

\revone{
We develop a sequential DA method to predict in real-time complex dynamical systems using low-order models.  
We define a low-order model as a qualitatively accurate model whose simulation time is of the same order as the timescales of experimental measurements (i.e., they provide real-time predictions), and whose  approximations and physical assumptions negatively affect its quantitative accuracy (i.e., low-order models may incur model bias).
}
The model bias must be treated explicitly to mitigate sub-optimal analysis~\citep{dee_bias_2005}. In a bias-aware framework, we need to provide an estimate of the bias at the time of assimilation, such that the biased forecast is mapped into an unbiased forecast, which, subsequently, can be combined with the observations data. 
This process is illustrated in Fig.~\ref{fig:tubes_BADA}.

Current bias-aware methods rely on defining \textit{a priori} a parameterized model to estimate the bias in the assimilation~\citep[e.g.,][]{auligne_adaptive_2007}. The parameters of the bias model are computed within the ensemble assimilation framework as a parameter estimation problem for $\vect{\beta}_j$. Thus, the cost function to minimize includes the estimation of the bias model parameters as

\begin{align}\label{eq:BA_cost_func_BB}
\mathcal{J}(\vect{\psi}_j, \vect{\beta}_j) = &\left\|\vect{\psi}_j-\vect{\psi}_j^\mtxt{f}\right\|^2_{\matr{C}^{\mtxt{f}^{-1}}_{\psi\psi}} +
 \left\|{\vect{y}}_j-\vect{d}_j\right\|^2_{\matr{C}^{-1}_{dd}}+
\left\|\vect{\beta}_j-\vect{\beta}_j^\mtxt{f}\right\|^2_{{\matr{C}^\mtxt{f}}^{-1}_{\beta \beta}}, \quad j=0,\dots,m-1, 
\end{align}
where $\vect{y}_j=\matr{M}\vect{\psi}_j + \vect{b}_j(\vect{\beta}_j)$ is the unbiased estimate,  
 % $\vect{b}_j$ is the bias of each ensemble member
 % (which differs from the expectation definition in~\S~\ref{sec:ensemble_framework} due to the ensemble of bias parameters  $\vect{\beta}_j$), 
and $\matr{C}^\mtxt{f}_{\beta \beta}$ is the bias parameters' error covariance matrix. 
\revtwo{This method is typically implemented in a variational approach~\citep[e.g.][]{tremolet_accounting_2006}. 
Alternatively, the parameters can be inferred through a sequential method using ensemble Kalman filters~\citep{dasilva_flow_2020}, which consist of augmenting the state vector with the bias parameters.}  
% }

The accuracy of the solution from~\eqref{eq:BA_cost_func_BB} depends on the suitability and complexity of the pre-defined functional form for the bias, $\vect{b}_j(\vect{\beta}_j)$. Increased complexity implies an increased number of model bias parameters to infer. 
However, augmenting the parameter estimation variables can  affect the conditioning of the minimization problem~\citep{dee_bias_2005}.
To overcome this limitation, \citet{novoa_magri_2022} proposed an Echo State Network (ESN), which is a reservoir computer tool, to address the challenge of bias estimation.  
The ESN bypasses the need of inferring the bias model parameters by autonomously inferring the bias in real time. Therefore, in contrast to the cost function dependent on bias parameters~\eqref{eq:BA_cost_func_BB},  \citet{novoa_magri_2022} use a sequential ensemble square-root Kalman filter that minimizes
\begin{align}\label{eq:BA_cost_func_JFM}
\mathcal{J}(\vect{\psi}_j) = &\left\|\vect{\psi}_j-\vect{\psi}_j^\mtxt{f}\right\|^2_{\matr{C}^{\mtxt{f}^{-1}}_{\psi\psi}} +
 \left\|{\vect{y}}_j-\vect{d}_j\right\|^2_{\matr{C}^{-1}_{dd}}, \quad j=0,\dots,m-1, 
\end{align}
where $\vect{y}_j=\matr{M}\vect{\psi}_j + \vect{b}^\mtxt{f}$, and 
the ESN estimates the bias forecast $\vect{b}^\mtxt{f}$, which is defined as an ensemble average (see~\S~\ref{sec:bf_details}). The network is re-initialized with the mean analysis innovation, $\vect{d}-\matr{M}\overbar{\vect{\psi}}^\mtxt{a}$, which allows the reservoir to be synchronized with the physical forecast model. We detail this implementation of the ESN in~\S~\ref{sec:ESN}.

\begin{figure}[!htb]
    \centering
    \includegraphics[width=.9\textwidth]{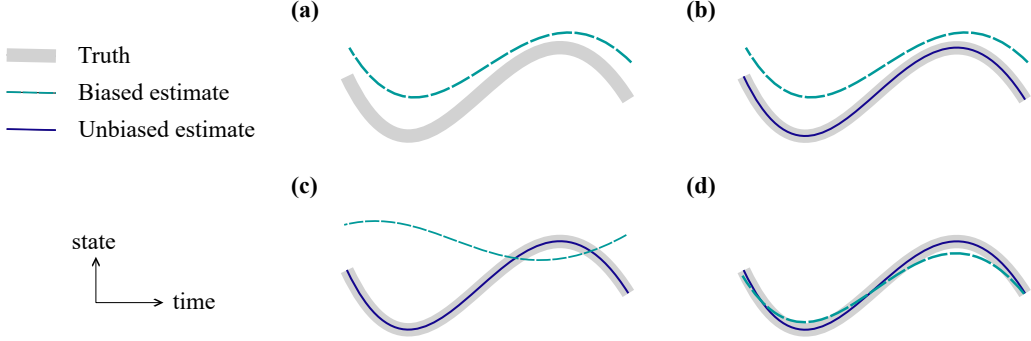}
    \caption{Pictorial comparison between 
    (a) bias-unaware data assimilation; 
    (b-d) bias-aware assimilation, in which the unbiased estimate (navy) matches the truth (grey).
    If the bias-aware framework is not regularized, it may indistinguishably recover solutions with  
    (b) unchanged, 
    (c) increased, or
    (d) reduced bias norm, as compared to the bias-unaware method. 
    The bias is seen as the distance between the biased estimate (dashed blue) and truth.
    The proposed regularized framework favours the small-norm bias scenario (d) if the bias regularization factor, $\gamma>0$.
     }
    \label{fig:BIAS_options}
\end{figure}

The bias-aware analysis schemes \eqref{eq:BA_cost_func_BB} and \eqref{eq:BA_cost_func_JFM} are designed to correct the bias of the forecast state (or equivalently in the observations) and, therefore, they reduce the mean analysis increments. However, they do not necessarily improve the analysis solutions, i.e.,~the assimilation may lead to an analysis state with a larger-norm bias than that of the forecast state. 
Figure~\ref{fig:BIAS_options} illustrates the different unbiased analyses that may  arise using current bias-aware DA methods. 
This raises the conceptual question ``what is a `good' unbiased analysis?''. 
We generalize the implementation of the ESN as the bias estimator in~\S~\ref{sec:ESN}.

\section{Regularized bias-aware ensemble Kalman filter (r-EnKF)} \label{sec:r-EnKF}

In a scenario with a biased model,  we judge that an unbiased analysis is `good' if the unbiased state (i)  matches the truth, and (ii) has a small-norm bias, as illustrated in Fig.~\ref{fig:BIAS_options}{d}. 
We are assuming that the model captures the main physical mechanisms needed to provide an estimate that is compatible with experimental data, i.e.,~the model is qualitatively accurate. Further, we assume that we can find a combination of parameters in the model that minimizes the norm of the model bias.
With this in mind, we formulate a data assimilation problem that not only does it minimize the estimation uncertainty and its distance to the observables, but it also reduces the bias of the estimator.  
Mathematically, we pose the problem by regularizing~\eqref{eq:BA_cost_func_JFM} such that
\begin{align}\label{eq:BR_cost_func}
\mathcal{J}(\vect{\psi}_j) = &\left\|\vect{\psi}_j-\vect{\psi}_j^\mtxt{f}\right\|^2_{\matr{C}^{\mtxt{f}^{-1}}_{\psi\psi}} +
 \left\|{\vect{y}}_j-\vect{d}_j\right\|^2_{\matr{C}^{-1}_{dd}}+\gamma\left\|\vect{b}_j\right\|^2_{\matr{C}^{-1}_{bb}}, \quad \mtxt{for} \quad j=0,\dots,m-1
\end{align}
where the operator $\left\|\cdot\right\|^2_{\matr{C}^{-1}}$ is the $L_2$-norm weighted by the semi-positive definite matrix ${\matr{C}^{-1}}$, and $\gamma\geq0$ is a bias regularization factor  (note that $\gamma$ could be merged with $\matr{C}_{bb}$, but as detailed in \S~\ref{sec:implementation}, we set $\matr{C}_{bb}=\matr{C}_{dd}$ and use the hyperparameter $\gamma$ to calibrate the relative sizes of the likelihood and the bias).
The `good' unbiased analysis that we define with cost function~\eqref{eq:BR_cost_func} accounts for the norm of the bias, in contrast to the possible outputs from current bias-aware frameworks which may result in analysis with increased or unchanged model bias~(Figs.~\ref{fig:BIAS_options}{b,c}). \\

An analysis state $\vect{\psi}_j^\mtxt{a}$ minimizes the regularized bias-aware cost function~\eqref{eq:BR_cost_func} if, for each ensemble member $j=0,\dots,m-1$,
\begin{align}\label{eq:BA_1} 
\dfrac{1}{2}\dfrac{\totald\mathcal{J}}{\totald\vect{\psi}_j}\bigg|_{\vect{\psi}_j^\mtxt{a}}&= \matr{C}^{\mtxt{f}^{-1}}_{\psi\psi}\left(\vect{\psi}_j^\mtxt{a}-\vect{\psi}_j^\mtxt{f}\right) + \dfrac{\totald{\vect{y}}_j}{\totald\vect{\psi}_j}\bigg|\T_{\vect{\psi}_j^\mtxt{a}}\matr{C}^{-1}_{dd}\left(\vect{{y}}^\mtxt{a}_j-\vect{d}_j\right)+\gamma\dfrac{\totald\vect{b}_j}{\totald\vect{\psi}_j}\bigg|\T_{\vect{\psi}_j^\mtxt{a}} \matr{C}^{-1}_{bb}\vect{b}_j^\mtxt{a} = 0.
\end{align}
The hyperparameter $\gamma$ directly affects the gradient on the left-hand side, whence,  the adaptability of the analysis state to the magnitude of the bias. With small bias regularization factors, $\gamma\ll1$, the analysis will evolve slowly with the bias norm, while large values of $\gamma$ allow large instantaneous updates to the ensemble state and parameters.
To mathematically solve~\eqref{eq:BA_1}, we assume that the analysis state is sufficiently close to the forecast such that we linearize the analysis bias as 
\begin{equation}\label{eq:linearization}
    \vect{b}_j^\mtxt{a} \approx \vect{b}_j^\mtxt{f} + \matr{J}^\mtxt{f}\matr{M}\left(\vect{\psi}_j^\mtxt{a} - \vect{\psi}_j^\mtxt{f}\right),
\end{equation}
where $\matr{J}^\mtxt{f}=\frac{\totald \vect{b}_j^\mtxt{f}}{\totald\matr{M}\vect{\psi}_j^\mtxt{f}}$ is the Jacobian. Further details on the forecast bias $\vect{b}_j^\mtxt{f}$ are discussed in~\S~\ref{sec:bf_details}. 
Importantly, the linearization~\eqref{eq:linearization} makes the  minimization problem quadratic.
Introducing~\eqref{eq:linearization} into \eqref{eq:BA_1}, grouping the terms in $\vect{\psi}_j^\mtxt{a}$, and exploiting the Woodbury matrix inversion formula~\citep{golub2013matrix}, yield the regularized bias-aware ensemble Kalman filter (r-EnKF)\footnote{The r-EnKF equations in the published version of this work~\citep{novoa2024inferring} have a typo. Equations \eqref{eq:r-EnKF} are the corrected form of the r-EnKF. }
\begin{equation}\label{eq:r-EnKF}
    \vect{\psi}_j^\mtxt{a} = 
    \vect{\psi}_j^\mtxt{f} + 
    \matr{K} \left[\left(\mathbb{I}+ \matr{J}^\mtxt{f}\right)\left(\vect{d}_j - \vect{y}_j^\mtxt{f}\right)\T - \gamma \matr{C}_{dd}\matr{C}^{-1}_{bb}{\matr{J}^\mtxt{f}}\T\vect{b}_j^\mtxt{f}\right],
\end{equation}
where the regularized Kalman gain matrix $\matr{K}$ is  
\begin{equation}\label{eq:r-EnKF_gain}
\matr{K} = \matr{C}_{\psi\psi}^\mtxt{f}\matr{M}\T\left[\matr{C}_{d d} + (\mathbb{I}+ \matr{J}^\mtxt{f})\T(\mathbb{I}+ \matr{J}^\mtxt{f})\matr{M}\matr{C}_{\psi\psi}^\mtxt{f}\matr{M}\T + \gamma \matr{C}_{dd}\matr{C}^{-1}_{bb}{\matr{J}^\mtxt{f}}\T\matr{J}^\mtxt{f}\matr{M}\matr{C}_{\psi\psi}^\mtxt{f}\matr{M}\T\right]^{-1}.
\end{equation}

\subsection{Discussion}
The r-EnKF equations~\eqref{eq:r-EnKF}-\eqref{eq:r-EnKF_gain} constitute a main result of this work as far as sequential data assimilation is concerned. 
We are defining a regularized bias-aware DA framework, which, in the limiting case of an unbiased (perfect) model, recovers a zero-bias state equivalent to the analysis from the bias-unaware EnKF~\eqref{eq:EnKF}. 
If the system is unbiased, i.e.,~$\vect{b}_j = \vect{0}$, the r-EnKF~\eqref{eq:r-EnKF} simplifies to the bias-unaware EnKF~\eqref{eq:EnKF}.
Furthermore, if the bias is independent of the state, i.e.,~$\matr{J}=\matr{0}$, the r-EnKF simplifies to the bias-aware EnKF resulting from~\eqref{eq:BA_cost_func_JFM}, in which the observations are assimilated into a bias-corrected forecast \citep[e.g.][]{novoa_magri_2022}.  

\revtwo{\citet{han_constrained_2016} proposed a cost function similar to~\eqref{eq:BR_cost_func}, and derived the constrained form of the variational bias-corrected 4D-Var. This variational method is restricted to pre-defined linear parametrized bias models, whose parameters are inferred in the assimilation analogously to~\eqref{eq:BA_cost_func_BB}.   
Including a pre-defined bias model in the assimilation can be beneficial if the form of the bias is known \textit{a priori}, and if it can be accurately represented with a pre-defined model. %However, if the bias is unknown, or if it is intricate (e.g., a nonlinear funciton of the state, or stochastic), implementing the method proposed by \citet{han_constrained_2016} in a sequential framework may not be possible. 
} 

\subsection{Simplification of the bias definition}\label{sec:bf_details}
\begin{figure}[!htb]
    \centering
    \includegraphics[width=\textwidth]{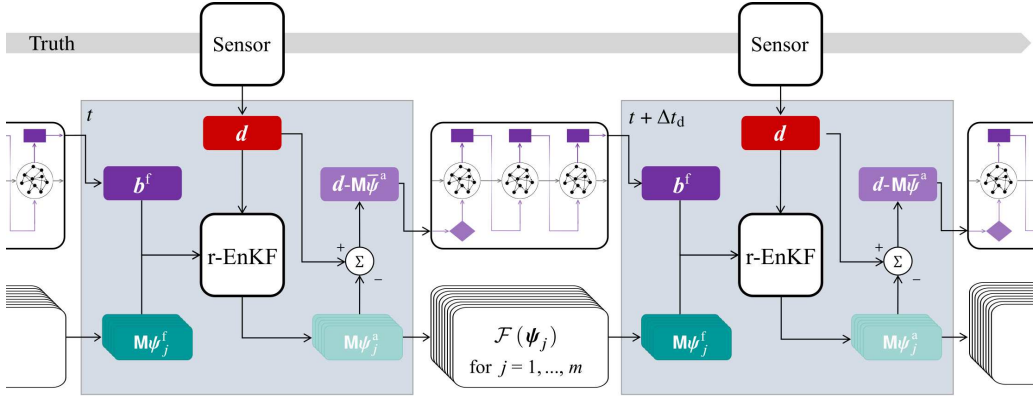}
    \caption{Schematic of the regularized bias-aware sequential data assimilation with an echo state network (ESN). 
    The low-order model $\mathcal{F}$ propagates each ensemble member to an ensemble of forecast states $\matr{M}\vect{\psi}_j^\mtxt{f}$, and the ESN runs autonomously in parallel to estimate the ensemble mean bias,  $\vect{b}^\mtxt{f}$.
    When a sensor provides noisy data $\vect{d}$, we apply the r-EnKF, which statistically combines the noisy data, the ensemble mean bias, and the forecast ensemble. The r-EnKF results in the analysis ensemble, $\matr{M}\vect{\psi}_j^\mtxt{a}$, which gives the new initial conditions for $\mathcal{F}$; and the analysis innovation $\vect{d}-\matr{M}\overbar{\vect{\psi}}^\mtxt{a}$ re-initializes the ESN.  
    This process is repeated sequentially every $\Delta t_\mtxt{d}$ time between observations. 
    }
    \label{fig:BADA_schematic}
\end{figure}

For brevity, instead of computing an ensemble of biases, $\vect{b}^\mtxt{f}_j$, we work with one echo state network to directly infer the mean bias, $\vect{b}^\mtxt{f}$. 
Therefore, the ESN forecasts autonomously the bias $\vect{b}^\mtxt{f}$ in time~\citep{novoa_magri_2022}. 
When observations become available, we update every ensemble member with $\vect{b}^\mtxt{f}_j\equiv\vect{b}^\mtxt{f}$, for $j=1,\dots,m$,  employing the r-EnKF~\eqref{eq:r-EnKF}; and finally re-initialize the ESN with the mean analysis innovation, $\vect{d}-\matr{M}\overbar{\vect{\psi}}^\mtxt{a}$. 
This procedure is illustrated in Figure~\ref{fig:BADA_schematic} and detailed in \S~\ref{sec:implementation}. 
Importantly, the transient innovations are not equivalent to the bias, because the model state and parameters are not yet converged.  Whereas, at statistical convergence, the ESN prediction and the innovations coincide  because of~\eqref{eq:expected_bias}, i.e.,~at convergence,
$\langle\vect{b}^\mtxt{f} - (\vect{d}-\matr{M}\vect{\psi}_j^\mtxt{f})\rangle$ is Gaussian noise, where $\langle\cdot\rangle$ indicates the time-average. 

%================================================================================================================================================================================% 
\section{Echo state network: an adaptive model for the bias} \label{sec:ESN}

Echo state networks (ESNs) are a type of recurrent neural networks, which are designed to learn temporal correlations in data through an internal high-dimensional state, the \textit{reservoir}~\citep{jaeger2004harnessing, maass2002real}.  
In previous work~\citep{novoa_magri_2022}, we showed that ESNs can be used in sequential DA to adaptively infer the bias as the model state and parameters converge in a sequential DA framework.  This is achieved by re-initializing the network with the analysis innovations (see Fig.~\ref{fig:ESNforecast}{b}). 
% 
%\lm{LM to LM: Many models can close the equations... We choose ESN because}
The motivation behind employing an ESN to model the bias of a low-order model in a sequential Kalman filter framework is three-fold: 
(i) ESNs  are generalized nonlinear auto-correlation functions~\citep[][p.306]{aggarwal2018neural}, which are universal approximators~\citep{GRIGORYEVA2018495}, this means that  they can accurately approximate time-varying functions (provided that they are trained on sufficient data);
(ii) the ESN can be formulated in the state space; and 
(iii) training an ESN consists of solving a linear-regression problem~\citep{lukovsevivcius_practical_2012}, which is time-efficient compared to backpropagation with gradient descent methods required in other recurrent architectures, such as long short-term memory networks~\citep{aggarwal2018neural}.

In the upcoming subsections, we summarize the state-space description of the ESN (\S~\ref{sec:ESN_state_space}), and the training process (\S~\ref{sec:trainESN}); then, we design a multi-parameter training data generation for a versatile and robust ESN (\S~\ref{sec:multi_train}), and we derive the Jacobian of the ESN, which is required in the r-EnKF (\S~\ref{sec:ESNJacobian}).

\subsection{State-space formulation}\label{sec:ESN_state_space} 
%------------------------------------------

The equations governing the time evolution of the bias forecast, $\vect{b}^\mtxt{f}$, given by an ESN are
\begin{align}\nonumber
\label{eq:ESN_OG}
        \vect{b}^\mtxt{f}(t_{k+1}) &= \matr{W}_{\mtxt{out}}\left[\vect{r}(t_{k+1}); 1\right]\\
        \vect{r}(t_{k+1}) &= \textrm{tanh}\left(\sigma_\mtxt{in}\matr{W}_{\mtxt{in}}\left[{\vect{i}}(t_k)\odot\vect{g}; \delta_\mtxt{r}\right]+
        \rho\matr{W}\vect{r}(t_{k})\right), 
\end{align}
where 
 $\vect{i}(t_k)$ is the input to the reservoir;
$\vect{r}\in\mathbb{R}^{N_{r}}$ and $\matr{W}\in\mathbb{R}^{N_{r}\times N_{r}}$ are the reservoir state and matrix, with a number of reservoir states (or neurons) $N_r\gg N_q$; 
\revtwo{the operator $\left[~;~\right]$ indicates horizontal concatenation;} 
$\matr{W}_{\mtxt{in}}\in\mathbb{R}^{N_{r}\times (N_q+1)}$ and $\matr{W}_{\mtxt{out}}\in\mathbb{R}^{ N_q\times(N_r+1)}$ are the input and output matrices, respectively;
$\sigma_\mtxt{in}$ is the input scaling;
$\rho$ is the spectral radius; 
$\delta_\mtxt{r}$ is a constant used for  breaking  the symmetry of the  ESN (we set $\delta_\mtxt{r}=0.1$)~\citep{huhn_learning_2020}; 
$\vect{g}=[g_1,\dots, g_{N_q}]$ are the input normalizing factors with $g_q^{-1} = \max{(\vect{b}_q^\mtxt{tr})}-\min{(\vect{b}_q^\mtxt{tr})}$, i.e.,~$g_q^{-1}$ is the range of $q^\mtxt{th}$ component of the training data $\vect{b}^\mtxt{tr}$ (the superscript `tr' stands for `training', \revtwo{which should not be confused with the superscript `t' for `true'}); 
the operator $\odot$ is the Hadamard product, i.e.,~component-wise multiplication; 
and the ${\tanh(\cdot)}$ operation is performed element-wise.  
The matrices $\matr{W}_{\mtxt{in}}$ and $\matr{W}$ are fixed, sparse and randomly generated using a connectivity of 5 \citep[see][for details]{jaeger2004harnessing, racca_robust_2021}, 
whereas $\matr{W}_{\mtxt{out}}$ is obtained through training, which is discussed in~\S~\ref{sec:trainESN}. 
The hyperparameters $\sigma_\mtxt{in}$ and $\rho$ are obtained with a recycle validation strategy and a Bayesian optimization following~\citep{racca_robust_2021}. 
The recycle validation is summarized in~\ref{app:ESN_RV} for completeness. %, and \ref{app:params} reports the ESN parameters used in the numerical test cases~\S~\ref{sec:test_cases}.

\begin{figure}[!t]
    \centering
    \includegraphics[width=.9\textwidth]{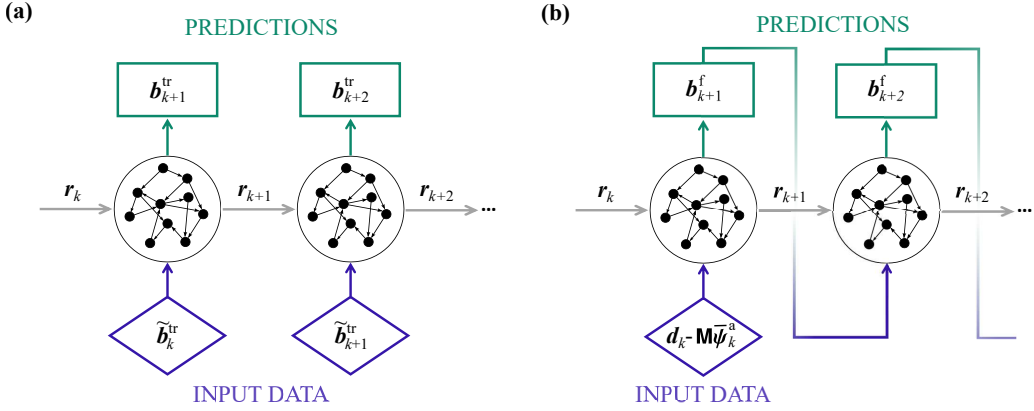}
    \caption{Schematic of the ESN forecast configurations during (a) training, and (b) assimilation. 
    The subscripts indicate time (e.g., $\vect{b}_k=\vect{b}(t_k)$).  
    During training, the ESN is forecast in open-loop, in which the input data (indicated with diamonds) are the noisy training data, and the training aims to match the output to the noise-free training data. 
    During the assimilation, the reservoir is re-initialized at the analysis step with the analysis innovation, and, from here, it runs autonomously in closed-loop.
    }
    \label{fig:ESNforecast}
\end{figure}

The ESN can evolve in either open or closed loops, in which the main differences affect the input to the reservoir, $\vect{i}(t_k)$ in~\eqref{eq:ESN_OG}, which in Figure~\ref{fig:ESNforecast} is illustrated with diamonds.  Figure~\ref{fig:ESNforecast}{a} shows the ESN open-loop configuration employed during training, in which the input is the training bias data, i.e.,~$\vect{i}=\tilde{\vect{b}}^\mtxt{tr}$ (the tilde indicates added noise, as detailed in~\S~\ref{sec:multi_train}). 
On the other hand,  Fig.~\ref{fig:ESNforecast}{b} shows that, during assimilation, the reservoir is re-initialized  at every analysis step with the analysis innovations (which is equivalent to a single open-loop step), i.e.,~$\vect{i}=\vect{d}-\matr{M}\overbar{\vect{\psi}}^\mtxt{a}$, and, from here, the ESN runs autonomously in closed-loop using the output from the previous step as the input. The implementation is further discussed in~\S~\ref{sec:implementation}.

\subsection{Training the echo state network} \label{sec:trainESN} %------------------------------------------
A trained ESN is the global minimizer of a quadratic optimization problem~\citep{lukovsevivcius_practical_2012}  
\begin{mini}|s|[0]
    {\footnotesize{\matr{W}_\mtxt{out}}}{\sum_{k=1}^{N_\mtxt{tr}-1}\left\|\matr{W}_{\mtxt{out}}\left[\vect{r}(t_k); 1\right] - \vect{b}^\mtxt{tr}(t_{k})\right\|_2^2 + \lambda \left\|\matr{W}_\mtxt{out}\right\|_2^2}
    {}
    {\label{eq:minESN}}{}
    \addConstraint{\vect{r}(t_k)}{= \textrm{tanh}\left(\sigma_\mtxt{in}\matr{W}_{\mtxt{in}}\left[\tilde{\vect{b}}^\mtxt{tr}(t_{k-1})\odot\vect{g}; \delta_\mtxt{r}\right]+
        \rho\matr{W}\vect{r}(t_{k-1})\right)}
    \addConstraint{\vect{r}(t_0)}{=\vect{0}},
\end{mini}
where $N_\mtxt{tr}$ is the length of the training dataset; 
 $\lambda$ is the Tikhonov regularization hyperparameter; 
 $\vect{b}^\mtxt{tr}(t_k)\in\mathbb{R}^{N_q}$ is the training data at time $t_k$;
 and $\tilde{\vect{b}}^\mtxt{tr}$ is the training data with added Gaussian noise, which regularizes the problem~\cite[e.g.,][]{racca_robust_2021}. The only trainable parameters are the components of the matrix $\matr{W}_{\mtxt{out}}$.

During training, the ESN is in open-loop configuration (Fig.~\ref{fig:ESNforecast}{a}), and  the aim is to match the outputs (rectangles in the figure) obtained from the input noisy training data $\tilde{\vect{b}}^\mtxt{tr}$ (diamonds) to the noise-free data $\vect{b}^\mtxt{tr}$ by modifying the weights in $\matr{W}_\mtxt{out}$.
Because the reservoir state $\vect{r}$ is not a function of $\matr{W}_\mtxt{out}$, the minimization problem \eqref{eq:minESN} consists of solving the linear system~\eqref{eq:RidgeReg} (ridge regression) 
\begin{equation}
\label{eq:RidgeReg}
    \left(\matr{R}\matr{R}\T + \lambda \mathbb{I}_{N_r+1}\right)\matr{W}_{\mtxt{out}}\T = \matr{R} \matr{B}\T,
\end{equation}
where 
$\matr{R} = \left[[\vect{r}(t_{1}); 1]\,|\dots|\,[\vect{r}(t_{N_\mtxt{tr}}); 1]\right]$  
and $\matr{B}=\left[\vect{b}^\mtxt{tr}(t_1)\,|\dots|\,\vect{b}^\mtxt{tr}(t_{N_\mtxt{tr}})\right]$. 
We regularize the ESN by adding Gaussian noise to each component of the input training data, i.e.,~ $\Tilde{\vect{B}}_q=\vect{B}_q+\mathcal{N}(\vect{0}, 0.03\mtxt{std}(\vect{B}_{q}))$, where $\mtxt{std}(\vect{B}_{q})$ is the  standard deviation in time of the $q^\mtxt{th}$ row of $\matr{B}$~\citep{racca_data_2022}. 
Because the input noise regularizes the ESN, we find  that optimizing the Tikhonov parameter does not impact significantly the ESN performance, thus, we set the Tikhonov parameter to $\lambda=10^{-16}$ for numerical stability. 

\subsection{Multi-parameter training dataset with data augmentation}\label{sec:multi_train} %------------------
The ESN training data must be carefully selected to obtain a robust estimator for the bias. 
\citet{novoa_magri_2022} suggested a single-parameter approach where the ESN is trained on a timeseries of synthetic biases, created from an educated  initial guess in the model state and parameters $\vect{\psi}^0=(\vect{\phi}^0, \vect{\alpha}^0)$ from prior knowledge. 
This method provides accurate predictions of the bias if the actual bias of the model is similar to the training time series. However, we found that the ESN may diverge if the ensemble is initialized far from $\vect{\psi}^0$, or if the dynamics of the innovations are significantly different to the single-parameter training dataset. 
To overcome this, we propose a multi-parameter training approach with data augmentation, which provides more physical information to the ESN. \\

The multi-parameter method consists of creating a set of $L$  timeseries of the bias data, whose initial guesses for the state and parameters are drawn from a uniform random distribution  $\mathcal{U}\left(\vect{\psi}^0(1-\sigma_L), \vect{\psi}^0(1+\sigma_L)\right)$, where  $\vect{\psi}^0$ is the mean, and $(1-\sigma_L, 1+\sigma_L)$ are the  normalized ranges of the uniform distribution.  
Then, we increase the number of training datasets by adding $2L$ synthetic bias timeseries with 10\% and 1\% amplitude of the original data (i.e., we add signals with physical dynamics but with smaller amplitudes). 
In the machine learning jargon, this is referred to as \textit{data augmentation}~\citep{goodfellow2016deep}. 
Data augmentation increases the robustness of the ESN for the network to infer the bias in lightly-biased models. (If the ESN is trained with large-norm bias datasets only, in the limiting case of an unbiased model, the ESN may not be able to recover a zero-norm bias.) 
The total number of training timeseries in the proposed training method is $3L$, thereby training the ESN consists of solving an augmented form of \eqref{eq:RidgeReg} such that 
\begin{equation}
\label{eq:RidgeReg_ens}
    \left(\sum_{l=0}^{3L-1}\matr{R}_l\matr{R}_l\T + \lambda\, \mathbb{I}_{N_r+1}\right)\matr{W}_{\mtxt{out}}\T = \sum_{l=0}^{3L-1}\matr{R}_l\matr{B}_l\T. 
\end{equation}
Although the computational cost of training the ESN increases with this approach, the summations in \eqref{eq:RidgeReg_ens} can be parallelized to minimize the impact of data augmentation in the training cost. 
\revtwo{Importantly, the ESN is not trained with the `true' bias, which is assumed to be \textit{a priori} unknown. Instead, we train the ESN with a diverse set of training bias datasets (created from an educated initial guess of the system's dynamics), which provides the ESN with a variety of physical information, and allows the network to reconstruct different biases as the assimilation converges. 
Thus, with the multi-parameter training we aim to alleviate the limited adaptivity of machine-learned bias models in sequential assimilation, which was discussed by, e.g., \citet{liang_machine_2023}.   
}

\subsection{Jacobian of the echo state network }\label{sec:ESNJacobian}%------------------------------------------
\toLM{
In this section, we derive the Jacobian required in the r-EnKF equations~\eqref{eq:r-EnKF}. For the ESN equations~\eqref{eq:ESN_OG}
\begin{equation}\label{eq:J_1}
    \matr{J} =\dfrac{\totald\vect{b}^\mtxt{f}(t_{k+1})}{\totald\matr{M}\vect{\psi}^\mtxt{f}(t_k)}= \dfrac{\totald\vect{b}^\mtxt{f}(t_{k+1})}{\totald\vect{i}(t_k)}\dfrac{\totald\vect{i}(t_{k})}{\totald\matr{M}\vect{\psi}^\mtxt{f}(t_k)}. 
\end{equation}
The first term in the right-hand side of~\eqref{eq:J_1} is the Jacobian of the ESN,  i.e.,~the gradient of the output with respect to the input; and the second  term in~\eqref{eq:J_1} is the gradient of the input with respect to the forecast biased model estimate. 
As shown in Fig.~\ref{fig:BADA_schematic}, we perform an open-loop step at the analysis step, such that the input to the ESN is the  mean analysis  innovation, i.e.,
\begin{equation}\nonumber
    \vect{i}(t_{k})=\vect{d}(t_{k})-\matr{M}\overbar{\vect{\psi}}^\mtxt{a}(t_{k})\quad \mtxt{and}\quad \matr{J} =-\dfrac{\totald\vect{b}^\mtxt{f}(t_{k+1})}{\totald\vect{i}(t_k)}. 
\end{equation}
In contrast, the ESN runs autonomously in between observations, in which  the input to the reservoir is the ESN prediction at the previous time step (see~ Fig.~\ref{fig:ESNforecast}{b}), thus, the second term in~\eqref{eq:J_1} is zero, and  $\matr{J}=\matr{0}$ in closed-loop.  
Hence, we only require the open-loop ESN Jacobian for the r-EnKF. (\ref{app:ESN_Jac}  includes the derivation of the closed-loop form for completeness.) }
\newline

\toLM{
To derive the Jacobian of the ESN, we split   the input and output matrices into
$   \matr{W}_\mtxt{in}=\left[\matr{W}_\mtxt{in}^{(1)}~\big|~\vect{W}_\mtxt{in}^{(2)}\right]
$ and 
$
\matr{W}_\mtxt{out}=\left[\matr{W}_\mtxt{out}^{(1)}~\big|~\vect{W}_\mtxt{out}^{(2)}\right],
$
where $\vect{W}_\mtxt{in}^{(2)}$ and  $\vect{W}_\mtxt{out}^{(2)}$ are the last column vectors, which only operate on the constants $1$ and $\delta_\mtxt{r}$, respectively. 
With this, we can rewrite~\eqref{eq:ESN_OG} such that the prediction at $t_{k+1}$ is a function of the quantities at $t_k$ only, i.e.,~
\begin{align}\label{eq:ESN_2}
        \vect{b}^\mtxt{f}(t_{k+1}) = \matr{W}_\mtxt{out}^{(1)}\tanh\left(\sigma_\mtxt{in}\matr{W}_\mtxt{in}^{(1)}\left(\vect{i}(t_k)\odot\vect{g}\right)+\sigma_\mtxt{in}\delta_\mtxt{r}\vect{W}_\mtxt{in}^{(2)} +\rho\matr{W}\vect{r}(t_k)\right)+ \vect{W}_\mtxt{out}^{(2)}.
\end{align}
In open-loop, the reservoir state is not a function of the input  because the input data is the innovation, rather than the prediction at the previous step. Therefore, at the analysis step (i.e., in open-loop), the  gradient of the output $\vect{b}^\mtxt{f}(t_{k+1})$ with respect to the input $\vect{i}(t_{k})$ yields
\begin{equation}\label{eq:J_open}
    \matr{J} =-\dfrac{\totald\vect{b}^\mtxt{f}(t_{k+1})}{\totald\vect{i}(t_k)} = -\matr{W}_\mtxt{out}^{(1)}\left[\matr{T}\odot\left(\sigma_\mtxt{in}\matr{W}_\mtxt{in}^{(1)}\odot\matr{G}\right)\right], 
\end{equation}
where $\matr{G}=\left[\vect{g}~|\dots|~\vect{g}\right]\T\in\mathbb{R}^{N_q\times N_r}$; 
$\vect{1}\in\mathbb{R}^{N_r}$ is a vector of ones; 
and  $\matr{T}=\left[\vect{T}~|\dots|~\vect{T}\right]\in\mathbb{R}^{N_r\times N_q}$ with
\begin{align}\nonumber
\vect{T}=\vect{1}-\tanh^2\left(\sigma_\mtxt{in}\matr{W}_\mtxt{in}^{(1)}\left(\vect{b}(t_k)\odot\vect{g}\right)+\sigma_\mtxt{in}\delta_\mtxt{r}\matr{W}_\mtxt{in}^{(2)} +\rho\matr{W}\vect{r}(t_k)\right). 
\end{align}
}

% ========================================================================================
\section{Regularized ensemble Kalman filter with an echo state network} \label{sec:implementation}

The implementation of the regularized bias-aware DA algorithm with an ESN can be divided into three steps: (1) ESN training, (2) ESN and ensemble initialization, and (3) the core of the assimilation. 
The first step is the training of the echo state network. We use a multi-parameter approach and data augmentation to generate the training set. The pseudo-algorithm~\ref{alg:train} summarizes this procedure, which is detailed in~\S~\ref{sec:trainESN}.
\begin{algorithm}[!htb]
\caption{Echo state network training}
\label{alg:train}
\begin{algorithmic}
\State ${\vect{d}}(t_0:t_\mtxt{tr})\gets$ \Call{GetObservations}{}
\For{$l=[0:L-1]$} \Comment{Multi-parameter data generation}
\State $\vect{\phi}_l, \vect{\alpha}_l = (\vect{\phi}^0, \vect{\alpha}^0) \mathcal{U}((1- \sigma_L), (1+ \sigma_L))$
\State $\matr{M}\vect{\psi}_l(t_0:t_\mtxt{tr}) \gets $  \Call{ForecastModel}{$\vect{\phi}_l, \vect{\alpha}_l$}
\State $\matr{B} \gets $ \Call{Append}{${\vect{d}} -\matr{M}\vect{\psi}_l$ }
\EndFor
\State $\matr{B} \gets $ \Call{Concatenate}{$\matr{B}, -0.1\matr{B}, 0.01\matr{B}$} \Comment{Augment training data}
\State $\matr{W}_\mtxt{out} \gets$ \Call{TrainESN}{$\matr{B}$}~\eqref{eq:RidgeReg_ens} 
\end{algorithmic}
\end{algorithm}
We need observation data sampled at a frequency $\Delta t_\mtxt{ESN}^{-1}$ for an interval $t_\mtxt{tr}$, which is problem-specific (refer to \ref{app:params} for the values of  the problem-specific parameters).  
We set the ESN time step larger than that of the physical model ($\Delta t_\mtxt{ESN}>\Delta t$), to reduce the computational cost added by the ESN. The time step should be, however, small enough to capture the dynamics of the bias.
%
% ----------------------------------------------------------------------------------------

The second stage shown in pseudo-algorithm~\ref{alg:washout} is the ensemble and ESN initialization.
\begin{algorithm}[!hbt]
\caption{Ensemble and echo state network initialization}
\label{alg:washout}
\begin{algorithmic}
\State $\vect{\psi}_j^\mtxt{f}(t_0)\gets $ \Call{InitializeEnsemble}{$m, \mathcal{N}(\vect{\alpha}^0, \sigma_\alpha\mathbb{I}), \mathcal{N}(\vect{\phi}^0, \sigma_\phi\mathbb{I})$}
\State{$t = t_\mtxt{startDA} - N_\mtxt{wash}\times\Delta t_\mtxt{ESN} - 2\Delta t_\mtxt{d}$}
\For{$j = 0:m-1$} \Comment{Parallel forecast}
\State{$\vect{\psi}_j^\mtxt{f}(t) \gets $  \Call{ForecastModel}{$\vect{\psi}_j^\mtxt{f}(t_0)$} \eqref{eq:ModeForecasrEnsemble}}
\EndFor
\Loop{ $N_\mtxt{wash}$ times:}\Comment{ESN washout}
\State $\matr{B}_\mtxt{wash} \gets$ \Call{Append}{$\vect{d}(t) - \matr{M}\overbar{\vect{\psi}}^\mtxt{f}(t)$ }
\For{$j = 0:m-1$} 
\State $\vect{\psi}_j^\mtxt{f}(t+\Delta t_\mtxt{ESN}) \gets $  \Call{ForecastModel}{$\vect{\psi}_j^\mtxt{f}(t)$} \eqref{eq:ModeForecasrEnsemble}
\EndFor
\State $t = t +  \Delta t_\mtxt{ESN}$
\EndLoop
\State $\vect{b}^\mtxt{f}(t), \vect{r}(t) \gets $ \Call{OpenLoopESN}{$\matr{B}_\mtxt{wash}$}  \eqref{eq:ESN_OG}%\Comment{Initialized ESN}
\end{algorithmic}
\end{algorithm}
As discussed in~\S~\ref{sec:BB}, the choice of the initial ensemble plays an important role on the assimilation because the analyses are confined to the range of the ensemble covariance, i.e.,~the analysis update results from a combination of the forecast ensemble members~\citep{evensen_ensemble_2003}.  The ensemble size and spread required for convergence depend on the dimensionality of the dynamical system at hand. We initialized the ensemble state and parameters from a multivariate Gaussian distribution as shown in the pseudo-algorithm~\ref{alg:washout}. 
The ensemble is forecast in time until statistical convergence. 
Then, we initialized the ESN with the mean innovations, which is also known as the \textit{washout} phase~\citep{lukovsevivcius_practical_2012}. The initialization must be performed before the start of the assimilation to update the reservoir state $\vect{r}$ with the physical bias. 
For a number of washout steps $N_\mtxt{wash}$ in open loop (Fig.~\ref{fig:ESNforecast}{a}), we feed input data from observations to the ESN every $\Delta t_{\mtxt{ESN}}$. The number of washout steps for an accurate initialization of the reservoir is problem-specific. 
%

% ---------------------------------------------------------------------------------------------

\begin{algorithm}[!htb]
\caption{Regularized bias-aware data assimilation with an echo state network}
\label{alg:BADA}
\begin{algorithmic}
\For{every observation $\vect{d}$ available}
\State ${\vect{d}}(t)\gets$ \Call{GetObservations}{}
\State $\matr{J} \gets $ $- $\Call{JacobianOpenLoop}{$\vect{b}^\mtxt{f}(t),\vect{r}(t)$} \eqref{eq:J_open}
\State $\vect{\psi}_j^\mtxt{a} \gets $ \Call{r-EnKF}{$\vect{\psi}_j^\mtxt{f}(t), \vect{b}^\mtxt{f}(t), \vect{d}(t),\matr{J}, \gamma, \matr{C}_{bb}$} \eqref{eq:r-EnKF} \Comment{Perform assimilation}
\If{all $\vect{\alpha}_j^\mtxt{a}$ are physical}{ \textbf{for} $j=0,\dots m-1$:}
\State {$\vect{\psi}_j^\mtxt{a} = \overbar{\vect{\psi}}^\mtxt{a} + 1.002~(\vect{\psi}_j^\mtxt{a} - \overbar{\vect{\psi}}^\mtxt{a})$}
\Else
\State{$\vect{\psi}_j^\mtxt{a} =\overbar{\vect{\psi}}^\mtxt{f}(t) + 1.05~(\vect{\psi}_j^\mtxt{f}(t) - \overbar{\vect{\psi}}^\mtxt{f}(t))$} 
\EndIf
\For{$j=0:m-1$}\Comment{Parallel forecast} 
\State{$\vect{\psi}_j^\mtxt{f}(t : t+\Delta t_\mtxt{d}) \gets $  \Call{ForecastEnsemble}{$\vect{\psi}_j^\mtxt{a}$}} \eqref{eq:ModeForecasrEnsemble} 
\EndFor
\State $\vect{b}^\mtxt{f}(t : t+\Delta t_\mtxt{d}),\vect{r}(t+\Delta t_\mtxt{d}) \gets $  \Call{ClosedLoopESN}{$\vect{d}(t) - \matr{M}\overbar{\vect{\psi}}^\mtxt{a}, \vect{r}(t)$} \eqref{eq:ESN_OG}
\State $t = t + \Delta t_\mtxt{d}$
\EndFor
\end{algorithmic}
\end{algorithm} 
The final step is the core stage of the assimilation, which is illustrated in Figure~\ref{fig:BADA_schematic} and it is detailed in pseudo-algorithm~\ref{alg:BADA}. 
The forecast model and the ESN are iteratively evolved in between observations for a time $\Delta t_\mtxt{d}$. Every $\Delta t_\mtxt{d}$, we 
(i) compute the open-loop Jacobian of the ESN~\eqref{eq:J_open}; and 
(ii) perform the analysis step by combining the ensemble forecast, the bias estimate, the observations, and the Jacobian, as well as the bias regularization factor, $\gamma$, and matrix weighting the bias-norm $\matr{C}_{bb}$. 
Because we define the bias in the observable space (see~\S~\ref{sec:problem_statement}), the observation error covariance matrix $\matr{C}_{dd}$ is an appropriate weight in the bias norm. Thus, we set $\matr{C}_{bb} =\matr{C}_{dd}$.
Through this procedure, the r-EnKF~\eqref{eq:r-EnKF} obtains the optimal combination between the biased forecast, the bias estimate, and the observations. The analysis indirectly updates the biased ensemble (as illustrated in Fig.~\ref{fig:tubes_BADA}). 
Because sequential filters do not enforce physical boundaries on the estimated parameters, we apply the reject-inflate strategy proposed by \citet{novoa_magri_2022}.  
If the resulting parameters are physical, we keep the analysis and apply a multiplicative inflation to the ensemble, motivated by \citep{da_ensemble_2018}.  
Alternatively, if the recovered parameters are not physical (e.g. positive time delays), we reject the analysis, and the algorithm returns the forecast ensemble with a stronger inflation with a 1.05 factor.
The inflation consists of multiplying the ensemble covariance by an inflation factor close to 1 (1.002 in our case~\citep{novoa_magri_2022}). This is common practice when using small ensemble sizes, which can systematically underestimate the error covariances, thereby potentially leading to  covariance collapse~\citep{van_comment_1999, evensen_data_2009}.
Finally, the analysis innovation, i.e.,~ $\vect{d}- \matr{M}\overbar{\vect{\psi}}^\mtxt{a}$ is used to re-initialize the ESN with a single open-loop step as illustrated in~Fig~\ref{fig:BADA_schematic}. The ESN is then evolved in time in closed-loop in parallel to the ensemble forecast, in which each ensemble member is evolved individually (as indicated by the stacking of $m$ models in Fig.~\ref{fig:BADA_schematic}).
% 
 
% ========================================================================================
% ========================================================================================
\section{Numerical test cases} \label{sec:test_cases}

%The proposed framework is versatile and can be applied to any time-varying numerical model.  
In this work, we showcase the r-EnKF to predict dynamics given by nonlinear oscillators.
Nonlinear oscillators are employed in variety of disciplines, ranging from quantum physics~\citep{dykman_quantum_2012} to cardiovascular studies~\citep{drzewiecki_bio_2021},   
and the synchronization property of coupled oscillators is exploited in a variety of  applications such as in laser physics,  to increase power output~\citep[e.g.][]{roy_laser_1994}, or in  nano-mechanics, to remove frequency differences arising from imperfections in fabrication~\citep{cross_synchronization_2006}. 
Further, nonlinear oscillators are extensively used to model vibrations and acoustics~\citep[e.g.][]{nguyen2020vibration}. 
Within the field of acoustics, thermoacoustic oscillations, which can also be modelled with nonlinearly coupled oscillators, are a key phenomenon in the fields of aviation and power generation  \citep{rayleigh_explanation_1878, lieuwen_book_2012, magri_adjoint_2019, sujith_pof_2020, magri_ARFM_2023}. 
Thermoacoustics are high-amplitude oscillations that arise when an unsteady heat source interacts with the  acoustics and hydrodynamics of a system. If the acoustic waves and the heat release are sufficiently in phase, the acoustic energy in the system increases, thus, a thermoacoustic oscillation arises. We showcase the r-EnKF on problems that are relevant to thermoacoustics because they offer computational challenges, such as nonlinearities and time-delays, which are common to a variety of nonlinear systems in engineering and science. 

We test the r-EnKF algorithm on two prototypical systems that have been employed in fundamental studies of thermoacoustic instabilities: (1) the van der Pol oscillator, which is a nonlinearly-forced oscillator~\citep{vanderPol_1926};  and (2) a nonlinearly time-delayed model, which is a set of nonlinearly coupled oscillators~\citep[e.g.][]{dowling_nonlinear_1997}. 
The problem-specific parameters of the ESN and DA framework, as well as the model parameters, are reported in~\ref{app:params}.

\subsection{Error metrics}
We use two error metrics to analyse the results.
First, we compute the time-evolution of the mean absolute error (MAE) of the biased and unbiased signals, relative to the maximum amplitude of the truth. For two quantities $\vect{w},\vect{z}$ with $N_q$ dimensions, the $\mtxt{MAE}$ is defined as
\begin{align}
    \mtxt{MAE}(\vect{w}, \vect{z}) &= \dfrac{1}{N_\mtxt{err}}\sum_{k=0}^{N_\mtxt{err}-1}\sum_{q=0}^{N_q-1} \dfrac{\left|\vect{w}_q({t_k})- \vect{z}_q({t_k})\right|}{\max{(\vect{d}_q^\mtxt{t})}}. 
\end{align}
The MAE is computed at every time interval $t_\mtxt{err}$ (which is problem-specific, see~\ref{app:params}) to recover a smooth time-evolution of the metric, that is, we analyse the moving-average of the absolute error. 
Second, we calculate the normalized root mean square error $(\mtxt{RMS})$ as
\begin{align}
    \mtxt{RMS}(\vect{w}, \vect{z}) &= \sqrt{\dfrac{\sum_{q}\sum_{k} \left(\vect{w}_{q}({t_k})- \vect{z}_{q}({t_k})\right)^2}{\sum_{q} \sum_{k}\left(\vect{w}_{q}({t_k})\right)^2}}, \quad \mtxt{for} \;
    \begin{array}{l}
    q=0,\dots, N_q-1, \\[.5em]
    k=0,\dots,N_\mtxt{err}-1
    \end{array}
\end{align}
% The RMS is computed in three time windows: 
% (i) $t_\mtxt{err}$ before the start of the assimilation (), 
% (ii)  the last $t_\mtxt{err}$ of the assimilation (DA), and 
% (iii) a time $t_\mtxt{err}$ after removing the filtering (post-DA).
% 
Table~\ref{tab:errors_names} summarizes the error metrics and their corresponding terminologies, which  we use in~\S~\ref{sec:VdP} and~\ref{sec:Rijke} to analyse the results. 
\begin{table}[!htb]
\caption{Terminology of the RMS and MAE metrics and their corresponding mathematical expression, and the time-windows in which the errors are computed.}
\centering
\label{tab:errors_names}
\begin{tabular}{@{}ll||ll@{}}
\toprule 
Term & Metric & Term & Time-window \\ \midrule
Biased RMS & $\mtxt{RMS}(\vect{d}^\mtxt{t}, \matr{M}\vect{\psi})$ & pre-DA  & $t_\mtxt{err}$ before the assimilation  \\
Unbiased RMS & $\mtxt{RMS}(\vect{d}^\mtxt{t}, \vect{y})$ & DA & last $t_\mtxt{err}$ of the assimilation \\
True biased RMS & $\mtxt{RMS}(\vect{d}^\mtxt{t}, \matr{M}\vect{\psi}^\mtxt{t})$ & post-DA & $t_\mtxt{err}$ after removing the filter \\
Biased MAE & $\mtxt{MAE}(\vect{d}^\mtxt{t}, \matr{M}\vect{\psi})$ &  &  \\
Unbiased MAE & $\mtxt{MAE}(\vect{d}^\mtxt{t}, \vect{y})$ &  & \\ \bottomrule
\end{tabular}
\end{table}

% ------------------------------------------------------------------------
\subsection{Van der Pol oscillator}\label{sec:VdP}
The van der Pol model is a forced oscillator with linear damping, whose time evolution is governed by the second-order differential equation
\begin{align}\label{eq:VdPwave}
    \ddot{\eta} + \omega^2{\eta} =  \dot{q} - \zeta\dot{\eta}, 
\end{align}
where $\omega$ is the angular oscillating frequency,
$\zeta$ is the damping coefficient, 
and $\dot{q}$ is a forcing term. 
When applied to thermoacoustics, $\eta$ represents the acoustic velocity, and $\dot{q}$ is the heat release rate. \citet{noiray2017linear} proposed  a nonlinear arc-tangent heat release law to model a thermoacoustic limit cycle from  experimental data, $\dot{q} = \beta\dot{\eta}\left(1 - {\kappa\eta^2}{(\beta + \kappa\eta^2)^{-1}}\right)$, where $\kappa$ is the nonlinearity coefficient, and $\beta$ is the forcing strength. 
Using this heat release law, Eq.~\eqref{eq:VdPwave} can be written as a system of ordinary differential equations as
\begin{subequations}
\begin{align}
    \dot{\eta} &= \mu\\
    \dot{\mu} &= -\omega^2\eta + \mu \left(\beta - \zeta -\dfrac{\beta\kappa\eta^2}{\beta + {\kappa}\eta^2}\right).
\end{align}
\end{subequations}
In reference to~\S~\ref{sec:ensemble_framework}, the state variables of the system are $\vect{\phi} = [\mu; \eta]$ and the physical parameters to infer are $\vect{\alpha} = [\zeta; \beta; \kappa]$. Hence, the augmented state vector is $\vect{\psi} = [\mu; \eta; \zeta; \beta; \kappa; \eta]$, where the biased model prediction is the acoustic velocity, $\matr{M}\vect{\psi} = \eta$. 

To showcase the r-EnKF, we create the reference data, i.e.,~the \textit{truth}, by prescribing a nonlinear additive bias to the observables created from a set of parameters $\vect{\alpha}^\mtxt{t}$ (reported in \ref{app:params}). In this case, we define 
\begin{equation}\label{eq:VdP_truth}
    {d}^\mtxt{t} = \eta^\mtxt{t} + \cos{\eta^\mtxt{t}}. 
\end{equation}
Therefore, the `true' bias of the model is $b^\mtxt{t}=\cos{\eta^\mtxt{t}}$. 
\toLM{
The observations are then created from the truth by adding Gaussian  noise with standard deviation $\sigma_\mtxt{d}=0.01$, which is normalized by $\langle|{d^t}|\rangle$, i.e.,~the time-average of the absolute value of the true amplitude.}
\\

Further forms of bias and types of observation noise are studied in the time-delayed model in ~\S~\ref{sec:Rijke}, which is more general. We use the van der Pol model to (1) analyse the effect of the multi-parameter training with data augmentation (\S~\ref{sec:VdP_results_1}), and (2) to perform a parametric analysis on the bias regularization factor, $\gamma$ (\S~\ref{sec:VdP_results_2}).

\subsubsection{Analysis of the multi-parameter training approach with data augmentation}\label{sec:VdP_results_1}
% Why data augmentation? ----------------------------------------------------------------------
We start our analysis by addressing a question that we asked in this work: ``what is a `good' unbiased analysis?''. We do so by showing the effects of data augmentation and the multi-parameter training approach (see~\S~\ref{sec:trainESN}) on the biased and unbiased RMS in the van der Pol oscillator. 
\begin{figure}[!ht]
    \centering
    \includegraphics[width=.9\textwidth]{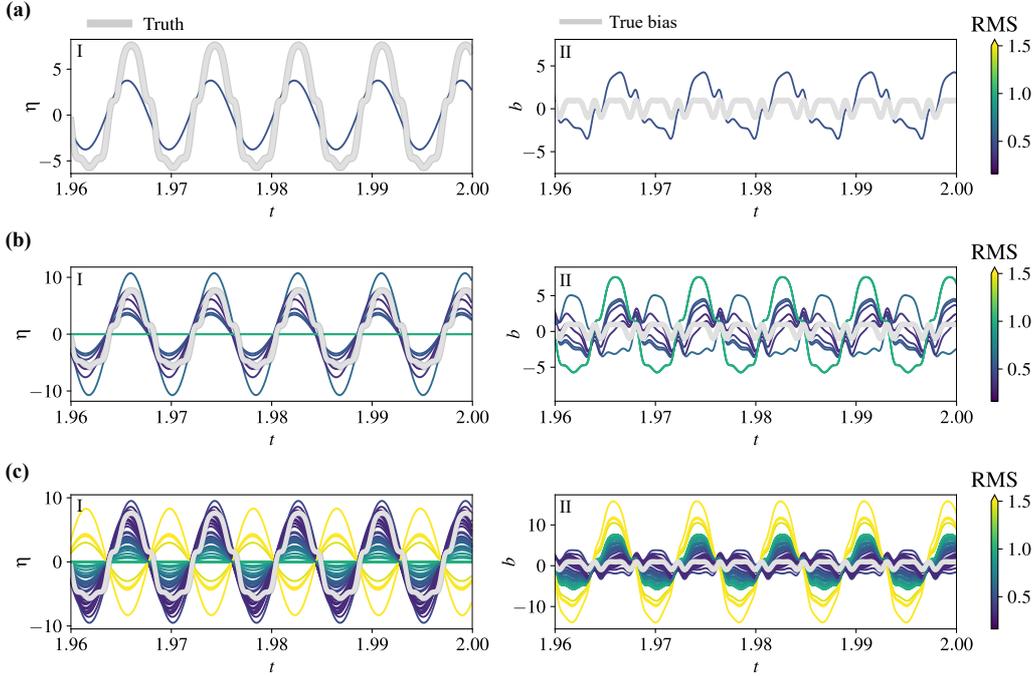}
    \caption{Van der Pol model. Comparison of the ESN training datasets with (a) single-parameter  training, i.e.,~$L=1$,  and multi-parameter training with (b) $L=10$, and (c) $L=50$. (I) Timeseries of the biased observables generated with a multi-parameter approach $\matr{M}\vect{\psi}_l = {\eta}_l$, for $l=1,\dots, L$ (colourmap), compared to the unbiased truth ${d}^\mtxt{t}$ (thick grey line). 
    (II) The corresponding $b_l$ training datasets for $l=1,\dots,L$ (colourmap), compared to the true bias $b^\mtxt{t}=\cos{\eta^\mtxt{t}}~$\eqref{eq:VdP_truth}, (thick grey line).
    The colourmap shows the biased RMS error $\mtxt{RMS}(\vect{d}^\mtxt{t}, \matr{M}\vect{\psi}_l)$ in the shown time-window.
    }
    \label{fig:training_sets}
\end{figure}
Figure~\ref{fig:training_sets} shows the truth and the generated multi-parameter observable timeseries for three different $L$-values (Fig.~\ref{fig:training_sets}I), as well as their corresponding bias datasets used to train the ESN in the simulations (Figs.~\ref{fig:training_sets}II). The bias training datasets are the innovations $b_l = {d}-\matr{M}\vect{\psi}_l$ for $l=1,\dots, L$ computed over a training time  $t_\mtxt{tr}=1.0$~s, sampled every $\Delta t_\mtxt{ESN}=5\E{-4}$~s. 
The multi-parameter training ($L>1$) cases shown in Fig.~\ref{fig:training_sets}{b,c} provide rich dynamical information to the ESN during training, which makes the network robust to sudden transient changes in the state and parameters during assimilation (see~\S~\ref{sec:VdP_results_2}). 
\begin{figure}[!ht]
    \centering
    \includegraphics[width=.75\textwidth]{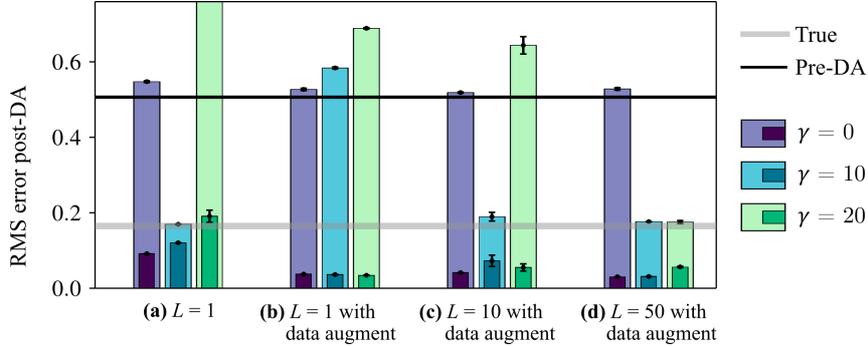}
    \caption{Van der Pol model. Analysis of the multi-parameter training with data augmentation, and the bias regularization hyperparameter. Comparison between the post-DA RMS errors using (a) single-parameter training, (b) single-parameter training with data augmentation, and  multiple-parameter training with data augmentation with (c) $L=10$ and (d) $L=50$;  
    with bias regularization factors $\gamma=0$ (violet), $\gamma=10$ (blue), and $\gamma=20$ (green). 
     The thicker bars show the biased error, $\mtxt{RMS}(\vect{d}^\mtxt{t}, \matr{M}\overbar{\vect{\psi}})$, and the overlapping thinner bars show the unbiased error, $\mtxt{RMS}(\vect{d}^\mtxt{t}, \overbar{\vect{y}})$. The error bars indicate the spread of the RMS errors in the ensemble. 
     The horizontal lines indicate the true biased RMS, $\mtxt{RMS}(\vect{d}^\mtxt{t}, \matr{M}\vect{\psi}^\mtxt{t})$ (grey), and the biased RMS before applying the filter (black). Solutions (c) with $\gamma=10$, and (d) with $\gamma=10, 20$ are `good' unbiased analysis because the unbiased RMS is small and the biased RMS is at true biased RMS levels.
     }
    \label{fig:WhyAugment}
\end{figure}
The effect of the multi-parameter training with data augmentation is analysed in Figure~\ref{fig:WhyAugment}, which shows the biased and unbiased RMS errors (see Tab.~\ref{tab:errors_names}) computed after the filter has been removed in four cases: a single-parameter training (a) without data augmentation, and (b) with data augmentation; and two multi-parameter training cases with data augmentation with (c) $L=10$, and (d) $L=50$. The $L$-datasets used for training the ESN correspond to those in Fig.~\ref{fig:training_sets}.
The initial ensemble is identical in each simulation, with an initial biased RMS of 0.51 (pre-DA black line in Fig.~\ref{fig:WhyAugment}). The grey line indicates the true biased RMS, as defined in Tab.~\ref{tab:errors_names}.
Figure~\ref{fig:WhyAugment} shows that if the bias is not penalized (violet bars in Fig.~\ref{fig:WhyAugment}), i.e.,~when $\gamma=0$ in~\eqref{eq:BR_cost_func}, the four cases provide solutions with small unbiased RMS (smaller than the true biased RMS); however, their biased RMS remains at a similar level to the pre-DA RMS. This undesired scenario, in which the bias norm does not decrease with the assimilation, is illustrated in Fig.~\ref{fig:BIAS_options}{b}. 
By increasing $\gamma$, we promote a small-norm bias analysis.  
Fig.~\ref{fig:WhyAugment}{a} shows that the simplest training approach recovers a small biased RMS at $\gamma=10$ (blue), however, the ESN does not recover accurately the bias of the solution (see the increased RMS of the thinner blue bar). Moreover, both the filter and ESN diverge if the bias regularization factor is increased to $\gamma=20$. 
Introducing data augmentation (Fig.~\ref{fig:WhyAugment}{b}), improves the robustness of the ESN, as seen by the small unbiased RMS; however, the recovered solutions in the three $\gamma$ cases have a larger-norm bias than the initial predictions, thus, the unbiased analysis is not `good' (this corresponds to the scenario depicted in Fig.~\ref{fig:BIAS_options}{c}).   
The `good' unbiased analysis (cf. Fig.~\ref{fig:BIAS_options}{d}) is recovered when we introduce the multi-parameter training. 
Comparing the multi-parameter training cases in Figs.~\ref{fig:training_sets}b,c, we see that increasing the size of the training dataset to $L=50$ improves the filter and ESN performance because the biased RMS are at true biased RMS levels, the unbiased RMS remains small, and the ensemble uncertainty (represented by the error bars) is small for both $\gamma={10~\mtxt{and}~20}$. 
The results improve with $L$ because more training data mean more information given to the ESN (as shown in Fig.~\ref{fig:training_sets}). 
Therefore, increasing the information fed to the ESN makes the network robust to sudden transient changes in the state and parameter during the assimilation. 
% 

% ------------------------------------------------------------------------------------------------------------------------------------------------------------------------------------------------------------------------------------------------------------------------------
\subsubsection{Effects of the training size and the bias regularization factor}\label{sec:VdP_results_2}
\begin{figure}[!h]
    \centering    
    \includegraphics[width=.9\textwidth]{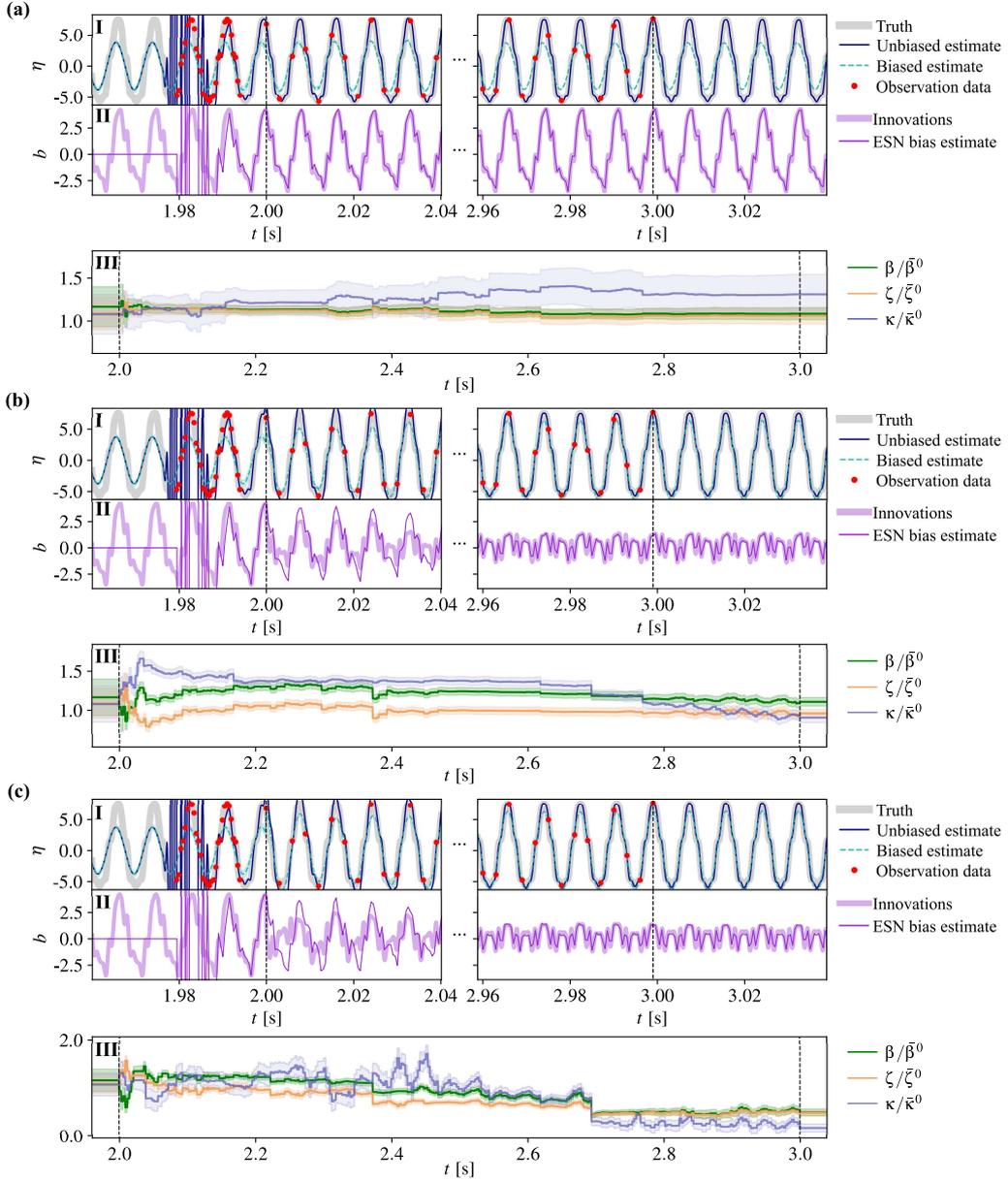}
    \caption{Van der Pol model. Analysis on the bias regularization factor in the assimilation, (a) $\gamma=0$, (b) $\gamma=10$, (c) $\gamma=20$, with $L=50$. 
       Time evolution at the start and end of assimilation of 
       (I) the true acoustic velocity (thick grey), unbiased estimate (navy), and biased estimate with its uncertainty (dashed teal); and 
       (II) the innovations (thick light orchid) and ESN prediction (thin dark orchid). 
       (III) Evolution of the inferred parameters and their uncertainty, normalized by their initial ensemble mean values.
       The assimilation window is indicated by the vertical dashed lines, and the red circles show the observation data.}
    \label{fig:VdP_time}
\end{figure}% 

Figure~\ref{fig:VdP_time} shows the effect of the bias regularization factor $\gamma$ on the time evolution of the simulations with $L=50$ in Fig.~\ref{fig:WhyAugment}. 
The bias and pressure timeseries (Figs.~\ref{fig:VdP_time}{I,II}) show that the ESN washout ($1.98~\mtxt{s}<t< 2$~s)  gives an accurate unbiased signal initialization (as seen by the unbiased estimate overlapping the truth, and the ESN estimate matching the innovations). 
 Consequently, if $\gamma=0$, the cost function~\eqref{eq:BA_cost_func_BB} is small regardless of the bias norm, and the innovations and parameters do not change significantly throughout assimilation. 
In contrast, if $\gamma>0$, the filter provides an analysis with a smaller-norm bias (as seen by the reduction in innovations, and the overlap of the biased estimate with the observations in Figs.~\ref{fig:VdP_time}{b,c}).  
As discussed in~\S~\ref{sec:r-EnKF}, $\gamma$ directly influences the adaptability of the analysis to the bias norm, which can be seen in Figs.~\ref{fig:VdP_time}{III}: the instantaneous changes in the model parameters at $\gamma = 20$ are larger than at $\gamma=10$. 
(For visualization, the parameters are normalized by their initial ensemble mean, which is not necessarily the solution to the problem, i.e.,~we do not expect the parameters to converge to 1.) However, the parameter convergence can be hindered if the instantaneous changes are too large.   
\begin{figure}[!htb]
    \centering
    \includegraphics[width=\textwidth]{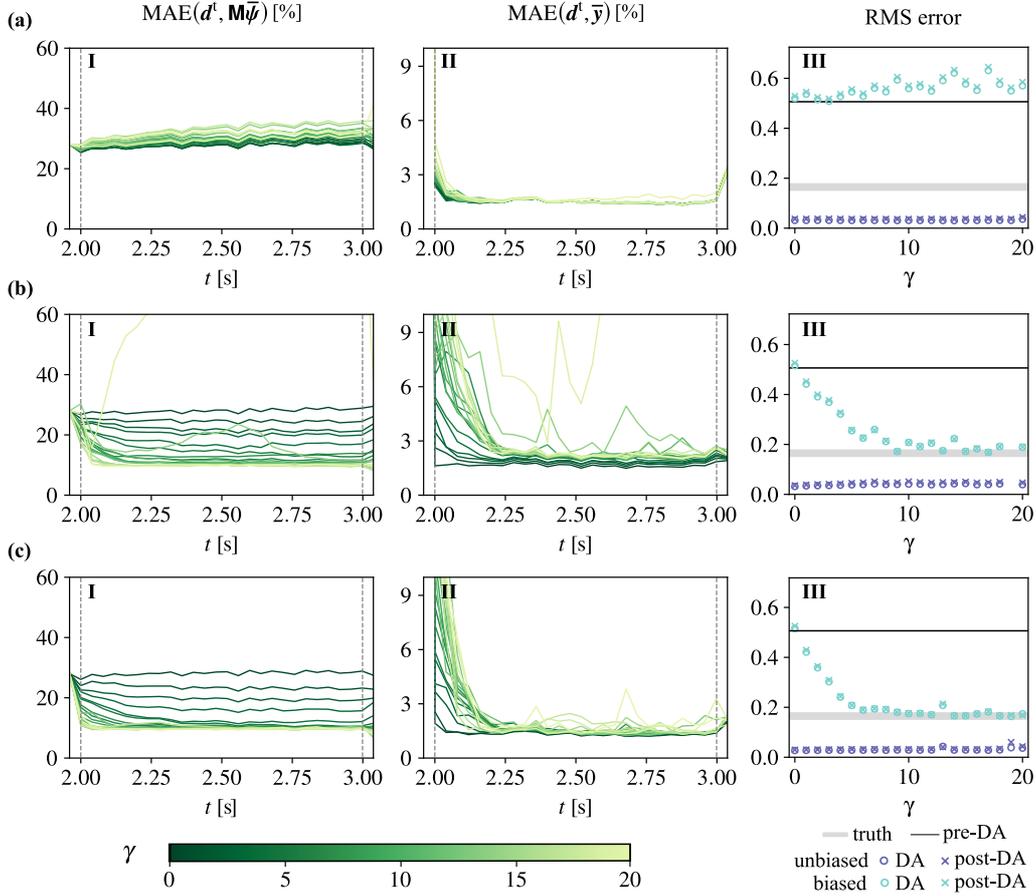}
    \caption{Van der Pol model. Parametric analysis of the bias regularization factor $\gamma$ for different numbers of training sets (a) $L=1$, (b) $L=10$, and (c) $L=50$. Evolution of the (I) biased and (II) unbiased MAE; and (III) RMS errors as defined in Tab.~\ref{tab:errors_names}: biased RMS (light blue) and unbiased RMS (dark blue) at convergence (circles) and after assimilation (crosses), true biased RMS (thick grey line), and biased RMS before assimilation (thin black line). 
    }
    \label{fig:errors_VdP}
\end{figure}
Therefore, tuning the bias regularization factor $\gamma$ is key to recovering small-norm bias analyses with the r-EnKF. 
We perform a parametric analysis on $\gamma$, for the 3 sizes of training datasets $L$.
Figure~\ref{fig:errors_VdP} shows the error metrics of the parametric analysis, as defined in Tab.~\ref{tab:errors_names}.  \revone{To complete the results of state, parameter and bias estimation, we include in \ref{app:params_convergence} the  convergence of the parameters for the same simulations.} 
With $L=1$, the filter performance is almost unaffected regardless of $\gamma$,  as shown by the small variation in the RMS and MAE throughout the assimilation in Fig.~\ref{fig:errors_VdP}{a}.  
Whereas with the designed multi-parameter training (Fig.~\ref{fig:errors_VdP}{b,c}), the filter is versatile and recovers `good' unbiased analysis, i.e.,~the the filter converges to solutions with biased RMS at true biased RSM levels, and  with small unbiased RMS values.   
Moreover, we see that at the ESN is more robust at $L=50$, because the prediction remains accurate for a wider range of $\gamma$,  and 
the unbiased RMS remains unchanged after the filter is removed. 
\revtwo{It is important to note that the ESN is only trained once for each of the $L$-values, i.e.,~for each row in Fig.~\ref{fig:errors_VdP}. Thus, even though the weights of the ESN are static, the network is adaptive enough to recover bias signals of different magnitudes and dynamics.}

Overall, the results shown in~\S~\ref{sec:VdP_results_1} and~\S~\ref{sec:VdP_results_2} indicate that 
(i) providing a rich variety of training datasets to the ESN (i.e. providing $L>1$) and performing data augmentation is essential to give the framework robustness and versatility; 
(ii) a proper tuning of the bias regularization factor is key to recovering a small-norm bias solutions with the r-EnKF; and 
(iii) the ESN is more robust and stable to wider ranges of the bias regularization factor as $L$ increases. 

\newpage

% ====================================================================================================================================
\subsection{Nonlinearly coupled time-delayed oscillators}\label{sec:Rijke}
In this section, we apply the r-EnKF to a set of coupled oscillators with a nonlinearly time-delayed forcing and nonlinear damping. We use a low-order time-delayed nonlinear thermoacoustic model, which is  widely used in fundamental studies of thermoacoustic oscillations~\citep[e.g.][]{heckl1990, balasubramanian_thermoacoustic_2008, magri_jfm_2013}. 
We formulate the time-delayed problem in the Markovian framework proposed by~\citet{huhn_stability_2019}. (The reader is referred to \ref{app:Rijke} for further details.) 
The discretized system of equations governing the acoustics is 
\begin{subequations}\label{eq:sys_modes}
\begin{align}
    &\dfrac{\mtxt{d}\eta_j}{\mtxt{d} t} = A_1\mu_j &\\
    &\dfrac{\mtxt{d} \mu_j}{\mtxt{d} t} = - A_2\eta_j - A_3\Dot{q}\sin{\left(A_4 x_\mtxt{h}\right)}- A_5\zeta_j \mu_j,   &\mtxt{for} \quad &j=0,\dots,N_m-1
    \\
    &\dfrac{\mtxt{d}\nu_i}{\mtxt{d}t} = 2\tau_\nu \sum_k\tensor{D}_{ik} \nu_k,    &\mtxt{for} \quad &i=0,\dots,N_c-1
\end{align}
\end{subequations}
where 
$\vect{\eta}$ and $\vect{\mu}$ are the acoustic velocity and pressure modes resulting from the Galerkin discretization of the acoustic velocity and pressure into $N_m$ acoustic modes; 
$\zeta_j$ is the damping, defined with the modal form $\zeta_j=C_1 j^2 + C_2 \sqrt{j}$~\citep{LANDAU_1987}; 
$\Dot{q}$ is the perturbation heat release rate, which is modelled with the time-delayed square-root model~\citep{heckl1990}
\begin{equation}\label{eq:qdot}
    \Dot{q}(t)= A_6\beta\left(\sqrt{\left|{\dfrac{1}{3}+A_7{u\left(x_\mtxt{h}, t-\tau\right)}}\right|}-\sqrt{\dfrac{{1}}{3}}\right), 
\end{equation}
with 
$\beta$ the heat release strength, and 
$\tau$ the time-delay such that $u\left(x_\mtxt{h}, t-\tau\right)$ is the acoustic velocity at the heat source location at the delayed time; 
and $\vect{\nu}$ are the spectral Chebyshev modes, which result from discretizing the advection equation modelling the history of $u\left(x_\mtxt{h}\right)$ into $N_c+1$ collocation points with the Chebyshev polynomial matrix  $\matr{D}$~\citep{trefethen_spectral_2000}; 
and the constants $A_1,\dots,A_7$ are specified in \ref{app:Rijke}. 
The advection equation transforms the time delayed problem into an initial value problem by
modelling a dummy variable $w$ travelling in a non-dimensional domain $X\in[0, 1]$ with velocity $\tau_\nu^{-1}$, with $\tau_\nu\geq\tau$. 
The Chebyshev modes are defined such that   
$\nu_0(t) = w(X=0,t) = u(x_\mtxt{h}, t)$, and $\nu_{N_c}(t) = w(X=1,t) = u(x_\mtxt{h}, t-\tau_\nu)$. 
Thus, the time-delayed velocity in \eqref{eq:qdot}, $u\left(x_\mtxt{h}, t-\tau\right)=w(X=\tau/\tau_\nu, t)$, which is computed by interpolation between Chebyshev modes.

The thermoacoustic parameters to infer are $\vect{\alpha} = [\beta; \tau]$; 
the state variables are $\vect{\phi} = [\vect{\eta}; \vect{\mu}; \vect{\nu}]$; and 
the observables in the nonlinear time-delayed system are acoustic pressure measurements at  $\vect{x}\in\mathbb{R}^{N_q}$ locations 
\begin{align}\label{eq:pressure}
    p_q(x_q,t)=-\sum^{N_m-1}_{j=0}\,\mu_j(t)\sin{\left(\dfrac{\omega_j}{\Bar{c}} x_q\right)}, \quad \mtxt{for} \quad q=0,\dots,N_q-1.
\end{align} 
With this, augmented the state vector is $\boldsymbol{\psi}=[\vect{\eta}; \vect{\mu}; \vect{\nu}; \beta; \tau; \vect{p}(\vect{x})]$, and the biased model prediction  is $\matr{M}\vect{\psi} = \vect{p}(\vect{x})$. 
We use $N_m=10$ for an appropriate representation of the acoustics~\citep{huhn_stability_2019}, and $N_q=6$ equidistant measurement locations, with $x_0=x_\mtxt{h}$ (i.e., the first measurement is at the heat source location), which is consistent with  experimental works~\citep{kabiraj_bifurcations_2011}. 
Because the time-delay $\tau$ is to be inferred and its value changes throughout the assimilation, 
we define $\tau_\nu=0.01$~s as the upper bound of the time-delay to give versatility to the filter~\citep[e.g.,][]{aguilar_sensitivity_2019}. We find that $N_c=50$ collocation points are required to avoid numerical dissipation in the advection equation.

\begin{figure}[!ht]
    \centering
    \includegraphics[width=\textwidth]{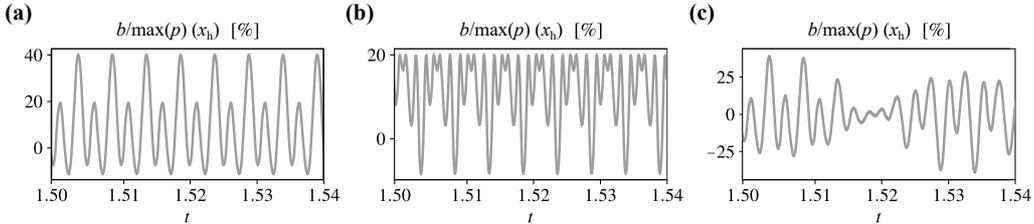}
    \caption{Nonlinear time-delayed model.  Synthetic bias signals at the heat source location ($x_\mtxt{h}=0.2$~m), normalized by the maximum amplitude of the true pressure at $x_\mtxt{h}$:
    (a) linear bias,
    (b) nonlinear bias, and
    (c) time-function bias.}
    \label{fig:rijke_biases}
\end{figure}

The true pressure is defined as $\vect{d}^\mtxt{t}=\vect{p}^\mtxt{t}+\vect{b}^\mtxt{t}$, where $\vect{b}^\mtxt{t}$ is the pre-defined bias. In this test case, we analyse three different forms of bias of increasing complexity: 
(i) a linear function of the acoustic pressure with an amplitude shift~\eqref{eq:Rijke_truth_linear},
which could correspond to an offset on the microphones calibration in a real experiment; 
(ii) a periodic transformation of the pressure~\eqref{eq:Rijke_truth_periodic}, which can model, for instance,  unclosed higher order acoustic modes (i.e., $N_m>10$); and  
(iii) an explicit time-dependent~\eqref{eq:Rijke_truth_time} bias which could arise in reality due to unmodelled physics in the low-order model (e.g., entropy waves, or mean flow effects), respectively 
\begin{subequations}
\begin{empheq}[left={\vect{b}^\mtxt{t} = \empheqlbrace\,}]{align}
&a_1\,{\vect{p}}^{\mtxt{t}}(\vect{x}) + a_2\max(\vect{p}^{\mtxt{t}}(x_\mtxt{h})). \label{eq:Rijke_truth_linear} \\ 
 &a_3\max(\vect{p}^{\mtxt{t}}(x_\mtxt{h})) \cos{\left(\frac{a_4\,{\vect{p}}^{\mtxt{t}}(\vect{x})}{\max(\vect{p}^{\mtxt{t}}(x_\mtxt{h}))}\right)}. \label{eq:Rijke_truth_periodic}\\
 &a_5\,{\vect{p}}^{\mtxt{t}}(\vect{x}) \sin{\left(a_6\pi t\right)^2}. \label{eq:Rijke_truth_time}
\end{empheq}
\end{subequations}
We set the constant values in~\eqref{fig:rijke_biases} to be $a_1=0.3, a_2=0.1, a_3=0.2, a_4=2, a_5=0.4$,  and $a_6=2$. 
Figure~\ref{fig:rijke_biases} shows the three types of bias in~\eqref{fig:rijke_biases} at the first measurement location ($x_\mtxt{h}$). 
In the linear case (Fig.~\ref{fig:rijke_biases}{a}) the bias retains the dynamics of the original signal (a 2-period limit cycle), but with a different mean and amplitude. 
The periodic dynamics of the original signal are also retained with the periodic bias forcing~\eqref{eq:Rijke_truth_periodic}, but with a different frequency spectrum (Fig.~\ref{fig:rijke_biases}{b}). 
In contrast, Fig.~\ref{fig:rijke_biases}{c} shows how the time-dependent bias~\eqref{eq:Rijke_truth_time} is not a limit cycle, but a quasi-periodic bias because, by adding a time-dependent sinusoidal, we are including a new incommensurate frequency to the dynamics.

\toLM{
Finally, to create the observations, we add to each component of the truth, Gaussian noise  with a standard deviation of $\sigma_\mtxt{d}=0.01$, which is normalized by the time-average of the mean absolute amplitude. 
We analyse the robustness of the r-EnKF for increasing values of $\sigma_\mtxt{d}$, as well as the impact of adding white, pink and brown noises of different strengths in \S~\ref{sec:noise}.  
}

% =================================================================================

\subsubsection{Analysis of  different biases: linear, nonlinear, and time-dependent} \label{sec:Rijke_results}
\begin{figure}[!htb]
    \centering
    \includegraphics[width=\textwidth]{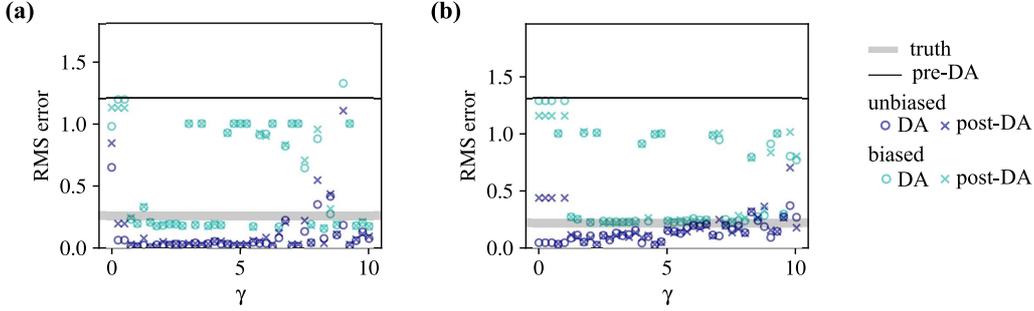}
    \caption{Nonlinear time-delayed model. RMS errors (as defined in Tab.~\ref{tab:errors_names}) with varying bias regularization factor $\gamma$, for the (a) linear and (b) nonlinear bias cases, with fixed $L=70$ and $m=50$. 
    RMS errors of the biased (light blue) and unbiased (dark blue) solutions at convergence (circles) and after assimilation (crosses), true biased RMS (thick grey line), and biased RMS of the initial ensemble (black line).}
    \label{fig:Rijke_CRP}
\end{figure}

In this section, we discuss the effect of the type of bias dynamics on the assimilation. We start from the linear and nonlinear bias cases. 
The number of degrees of freedom in the nonlinearly coupled oscillators is larger than in the van der Pol case. Thus, as discussed in~\S~\ref{sec:SDA}, the initialization of the ensemble (which is given by the ensemble size, $m$, and the initial ensemble spread) is key to the performance of the filter because the analyses are confined to the range of the forecast ensemble error covariance matrix. 
We perform the analysis shown in this section for three ensemble sizes. We find that $m=50$ provides a good trade-off between filter performance and computational cost (results shown in \ref{app:m10-80}).% 
\begin{figure}[!htb]
    \centering
    \includegraphics[width=\textwidth]{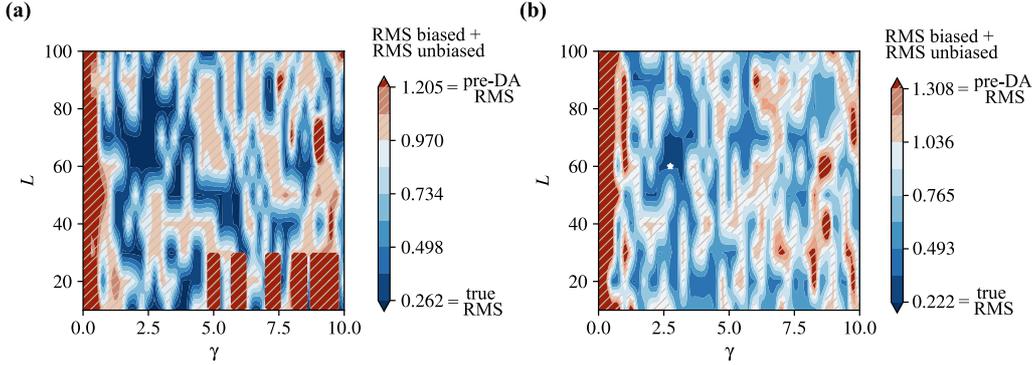}
    \caption{Nonlinear time-delayed model. Combined bias and unbiased RMS errors after assimilation with (a) linear, and (b) nonlinear biases. 
    Contours for varying training set size, $L$, and bias regularization factor, $\gamma$. 
    The colourmap shows the sum of the bias and unbiased RMS errors, and the minimum  combined RMS point is indicated with a white star.} 
    \label{fig:Contours_linear_periodic}
\end{figure}
Figure~\ref{fig:Contours_linear_periodic} shows, for the linear~\eqref{eq:Rijke_truth_linear} and nonlinear~\eqref{eq:Rijke_truth_periodic} biases, the sum of the biased and unbiased RMS (see Tab.~\ref{tab:errors_names}) after the filter is removed. 
The figure shows that bias regularization factors  $\gamma \gtrsim 5$ and $\gamma \lesssim 1$  tend to provide inaccurate solutions or even divergence. The areas in dashed dark red show solutions where the combined bias and unbiased RMS is larger than the true signal magnitude. 
The error is smaller in the linear bias case (Fig.~\ref{fig:Contours_linear_periodic}{a}), which is to be expected due to the increased difficulty arising from the nonlinear bias form (Fig.~\ref{fig:Contours_linear_periodic}{b}). 
We define optimal $(L^\star,\gamma^\star)$ pairs as the minimum values of the combined RMS error ${(\mtxt{RMS}_b + \mtxt{RMS}_u)}$ (i.e., the global minima in Fig.~\ref{fig:Rijke_CRP} are the post-DA $(L^\star, \gamma^\star)$). 
Table~\ref{tab:RMS} reports for each of the bias cases, the true biased RMS, and the biased and unbiased RMS of the optimal solutions at convergence, and after the filter is removed.
\begin{table}[!htb]
\caption{Nonlinear time-delayed model RMS errors, for the linear, nonlinear, and time-dependent bias cases.  Comparison between the true, $\mtxt{RMS}(\vect{d}^\mtxt{t}, \matr{M}\vect{\psi}^\mtxt{t})$, biased $\mtxt{RMS}(\vect{d}^\mtxt{t}, \matr{M}\overbar{\vect{\psi}})$, and unbiased $\mtxt{RMS}(\vect{d}^\mtxt{t}, \overbar{\vect{y}})$ at  assimilation convergence (DA) and after the filter is removed (post-DA).}
\centering
\label{tab:RMS}
\begin{tabular}{@{}lcrccrcc@{}}
\toprule
\multicolumn{1}{c}{\multirow{2}{*}{Bias type}} &
  \multirow{2}{*}{\makecell{True \\  biased  RMS}} &
  \multicolumn{3}{c}{{DA RMS}} &
  \multicolumn{3}{c}{post-DA RMS} \\ \cmidrule(l){6-8} \cmidrule(l){3-5} 
\multicolumn{1}{c}{} & &
  \multicolumn{1}{c}{$L^\star, \gamma^\star$} &
  \multicolumn{1}{c}{Biased} &
  \multicolumn{1}{c}{Unbiased} &
  \multicolumn{1}{c}{$L^\star, \gamma^\star$} &
  \multicolumn{1}{c}{Biased} &
  \multicolumn{1}{c}{Unbiased} \\ \midrule
Linear         &0.2623 & 10, 3.50 & 0.1761 & 0.0244 & 100, 1.75 &0.1817 & 0.0157 \\
Nonlinear      &0.2217 & 60, 2.75 & 0.2303 & 0.0799 & 60, 2.75& 0.2279 & 0.0792 \\
Time-func.  &0.2385 & 30, 0.50 &   0.2860   & 0.0590       & 10, 1.25 &   0.2534   &    0.4434    \\ \bottomrule
\end{tabular}
\end{table}

Figure~\ref{fig:Contours_linear_periodic} shows how for $L=70$ (i.e., training the ESN with 70 signals), the r-EnKF provides accurate solutions for a large range of $\gamma$ (as seen by the blue areas concentrated in this region). We show in Figure~\ref{fig:Rijke_CRP} the detailed RMS errors and the inferred parameters at  $L=70$, for the linear (Fig.~\ref{fig:Rijke_CRP}{a}) and nonlinear (Fig.~\ref{fig:Rijke_CRP}{b}) biases. 
The filter sensitivity to $\gamma$ is higher than that of the van der Pol model in~\S~\ref{sec:VdP_results_2}.  
In agreement with the van der Pol results, the biased RMS remains at pre-DA levels when the bias is not penalized, but as $\gamma$ increases, the biased RMS converges to true-RMS levels. 
Interestingly, Figure~\ref{fig:Rijke_CRP}{a} shows that the  r-EnKF finds an analysis solution with smaller-norm bias than the true linear bias, which is indicated by the biased post-DA RMS being smaller than the true biased RMS (see the exact values in  Table~\ref{tab:RMS}). 
Indeed, the objective of the proposed r-EnKF algorithm is to recover small-norm bias analyses, and, although we are prescribing a form of bias, this may not be the smallest-norm bias solution available. Thus, the r-EnKF can find a set of model parameters $\vect{\alpha}=[\beta; \tau]$ \toLM{(which are reported in \ref{app:params_convergence} for completeness)} that provides a smaller-norm bias than the true bias~\eqref{eq:Rijke_truth_linear}. 
% 
%This unique $\vect{\alpha}$ for minimum bias differs from the van der Pol model, which allows small bias signals for multiple combinations of model parameters.  
% 
Figure~\ref{fig:Rijke_CRP}{b} shows qualitatively similar results on the nonlinear bias case: consistent small-norm bias is recovered at $\gamma\in[1.5, 4.5]$, which are given by a unique $\vect{\alpha}$. 
The less accurate results are expected due to the increased difficulty in the problem (arising from the nonlinearity of the bias). 
The predictions remain accurate after the filter has been removed, which means that the filter recovers a `good' unbiased analysis, and the prediction window of the ESN is long in the linear and nonlinear cases. 
\begin{figure}[!htb]
    \centering
    \includegraphics[width=\textwidth]{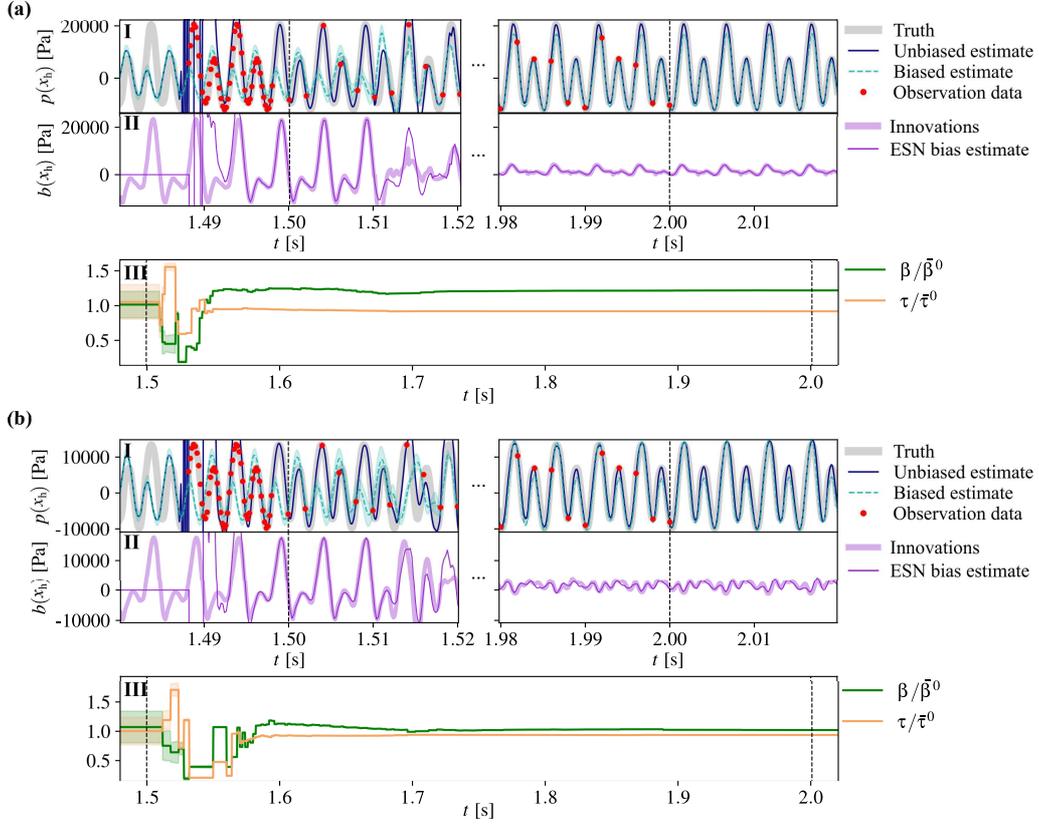}
    \caption{Nonlinear time-delayed model. Timeseries results with 
    (a) linear bias for $L^\star=100, \gamma^\star=1.75$; and 
    (b) nonlinear bias for $L^\star=60, \gamma^\star=2.75$. 
    Time evolution at the start and end of assimilation of 
    (I) the true acoustic velocity (thick grey), unbiased estimate (navy), and biased estimate with its uncertainty (dashed teal), and 
    (II) the innovations (thick light orchid) and ESN prediction (thin dark orchid); and 
    (III) time evolution throughout the assimilation of the inferred parameters and their uncertainty, normalized by their initial ensemble mean values. 
    The assimilation window is indicated by the vertical dashed lines, and the red circles show the observation data.}
    \label{fig:Rijke_time_1}
\end{figure}
Figure~\ref{fig:Rijke_time_1} shows the detailed timeseries  for the optimal $(L^\star, \gamma^\star)$ (see Tab.~\ref{tab:RMS}).  Figures~\ref{fig:Rijke_time_1}{I} compares at the start and end of the assimilation,  the true observable acoustic pressure at the heat source location, ${p}^\mtxt{t}(x_\mtxt{h})+{b}^\mtxt{t}(x_\mtxt{h})$, to the filter unbiased estimation, ${p}(x_\mtxt{h})+{b}(x_\mtxt{h})$, and the biased estimate, ${p}(x_\mtxt{h})$, which is the model estimate before applying the bias correction. 
The ESN bias estimate $b^\mtxt{f}(x_\mtxt{h})$ is compared to the true innovations (i.e., the difference between the truth and the biased estimate,  ${p}^\mtxt{t}(x_\mtxt{h})+{b}^\mtxt{t}(x_\mtxt{h}) - {p}(x_\mtxt{h})$). The agreement between the innovation and the ESN prediction shows that the ESN successfully infers the bias during the assimilation in open-closed loops, and the closed-loop prediction remains accurate after the assimilation ends. 
The evolution of the parameters throughout  the assimilations are shown in Figs.~\ref{fig:Rijke_time_1}{III}, which indicate that the filter converges rapidly to a combination of thermoacoustic parameters that provides the small-norm bias solution, i.e.,~with  significantly reduced innovations (as seen in Figs.~\ref{fig:Rijke_time_1}{II}). 
Importantly, the sampling frequency in real thermoacoustic oscillations experiments is typically 10~kHz~\citep[e.g.][]{kabiraj_nonlinear_2012}, which is higher than the sampling required in this framework. As reported in~\ref{app:params}, we use a model sampling frequency of 10~kHz, an ESN forecast frequency of 5~kHz, and the assimilation frequency is 0.5~kHz.

% ------------------------------------------------------------------------------------------------------------------------------------------------------

Finally, we address the case of the time-dependent bias defined in~\eqref{eq:Rijke_truth_time}. This is the most challenging scenario because the bias is not only a function of the state, but also an explicit  function of time. With the explicit time-dependency, we are 
 adding an incommensurate frequency to the dynamics, thereby creating a quasi-periodic bias. 
We   
(i) increase the ESN training time from 0.5~s to 1.5~s to allow the ESN to learn time-varying information (note that the ESN is not aware of the time information);  and we 
(ii) increase the assimilation frequency to 1~kHz to achieve accurate results (which is to be expected because quasi-periodic dynamics are more complex than limit cycles, and a higher sampling rate is required to reconstruct a quasi-periodic signal). 
\begin{figure}[!htb]
    \centering
    \includegraphics[width=.95\textwidth]{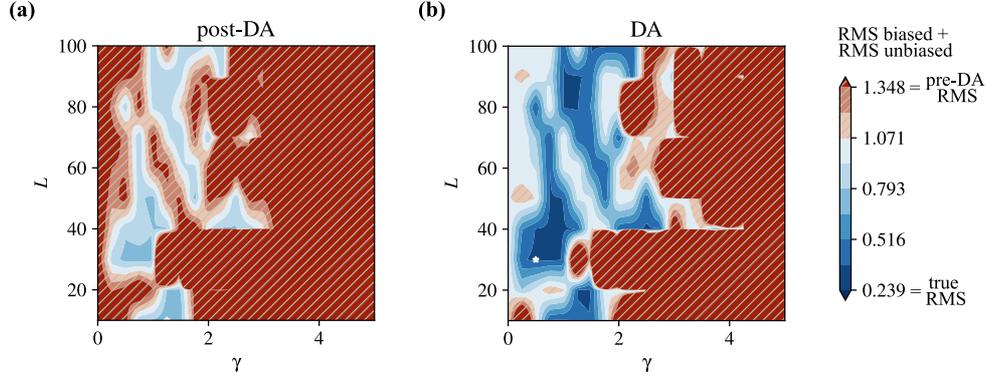}
    \caption{Nonlinear time-delayed model. Error contours for different $(L, \gamma)$ combinations for the explicit time-dependent bias. The colourmap shows the sum of the bias and unbiased RMS errors (a)  after the filter is removed (post-DA), and (b) at the end of the assimilation (DA). The minimum  combined RMS point is indicated with a white star.}
    \label{fig:Contours_time}
\end{figure}% 
Figures~\ref{fig:Contours_time}{a, b} show the combined RMS error contours after removing the filter and at the end of the assimilation, respectively. 
The allowed range of the bias regularization factor is visibly smaller than in the linear and nonlinear biases in Fig.~\ref{fig:Contours_linear_periodic}, and the combined RMS error is higher after assimilation. 
However, as shown by the DA RMS contours, the  error is small during assimilation, when both DA and post-DA biased RMS are of the same order of magnitude to the true biased RMS (see Tab.~\ref{tab:RMS}). 
This means that the filter successfully converges to a small-norm solution, but the ESN predictive window is shorter than in the linear and nonlinear baises due to the quasi-periodicity of the time-dependent bias. 
\begin{figure}[!htb]
    \centering
    \includegraphics[width=.95\textwidth]{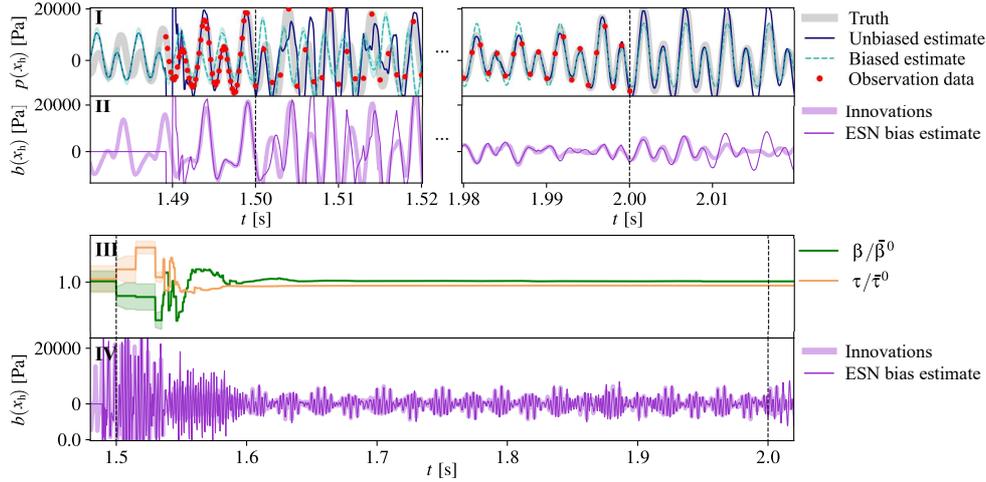}
    \caption{Nonlinear time-delayed model. Timeseries results with a time-dependent bias for $L^\star=10, \gamma^\star=1.25$. 
    Time evolution at the start and end of assimilation of 
    (I) the true acoustic velocity (thick grey), unbiased estimate (navy), and biased estimate with its uncertainty (dashed teal), and 
    (II) the innovations (thick light orchid) and ESN prediction (thin dark orchid); 
    and time evolution throughout the assimilation of 
    (III) the inferred parameters and their uncertainty, normalized by their initial ensemble mean values, and 
    (IV) the innovations and ESN bias prediction. 
    The assimilation window is indicated by the vertical dashed lines, and the red circles show the observation data.}   
      \label{fig:Rijke_time_2}
\end{figure}
We show the detailed timeseries results for the optimal post-DA $(L^\star, \gamma^\star)$ in Figure~\ref{fig:Rijke_time_2}. 
The biased solution remains close to the truth after the filter is removed (see Fig.~\ref{fig:Rijke_time_2}{I}). 
The ESN prediction becomes accurate with the re-initialization with the analysis innovations (as shown in  Fig.~\ref{fig:Rijke_time_2}{IV}). 
Assimilating the data with the unbiased signal allows the filter to perform an unbiased analysis, and the bias regularization guides the solution to a small-norm bias estimate (see~\S~\ref{sec:r-EnKF}). 

\subsubsection{Performance of the bias-unaware EnKF on the time-varying bias case}

In this section, we show how the traditional bias-unaware EnKF performs on a biased model. We show the time-varying bias scenario only for brevity. 
\begin{figure}[!htb]
    \centering
    \includegraphics[width=\textwidth]{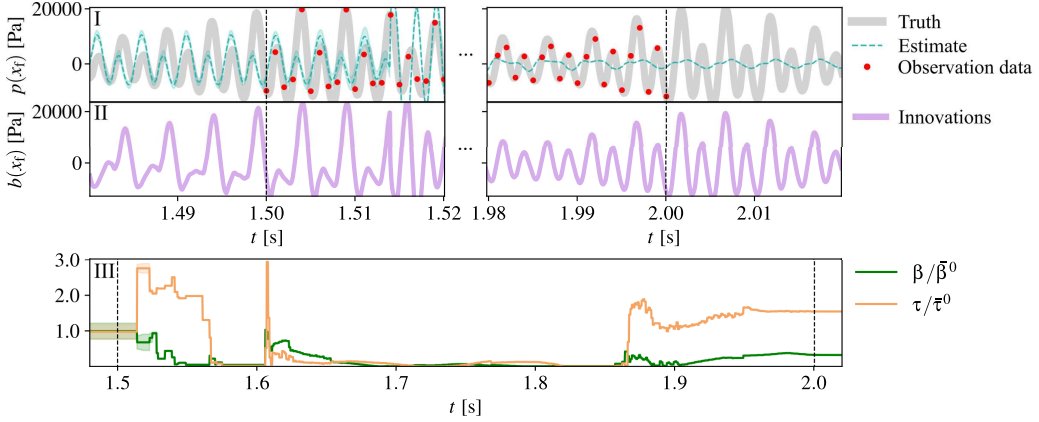}
    \caption{Nonlinar time-delayed model with a time-dependent bias. Results with a bias-unaware EnKF. 
    Time evolution at the start and end of assimilation of 
    (I) the true acoustic velocity (thick grey), pressure estimate with its uncertainty (dashed teal); and 
    (II) the innovations (thick light orchid). 
    Time evolution throughout the assimilation of 
    (III) the inferred parameters and their uncertainty, normalized by their initial ensemble mean values.
    The assimilation window is indicated by the vertical dashed lines, and the red circles show the observation data.}  
      \label{fig:Rijke_time_EnKF}
\end{figure}
The results in figure~\ref{fig:Rijke_time_EnKF} are the bias-unaware equivalent to Fig.~\ref{fig:Rijke_time_2}. Both simulations are performed using the same initial ensemble and filter parameters. 
Noticeably, the magnitude of the  innovations is large at the end of the assimilation, and the state estimate does not resemble the truth. Further, it can be seen that the bias-unaware filter attempts to converge, but at $t\approx 1.87$~s there is a sudden instantaneous update in the parameters. We expect this behaviour  to occur cyclically due to the time-varying bias. As discussed in~\S~\ref{sec:BB}, if unbiased data is assimilated into a biased model, the model drift can increase the bias of the analysis, because the bias-unaware framework assumes unbiased forecasts.

\subsubsection{Robustness of the r-EnKF to the level and colour of the observation noise}\label{sec:noise}
In this section, we analyse the impact of the strength and colour of the measurement noise on the assimilation. We perform the robustness study for the case of the nonlinear time-delayed model with nonlinear bias.

\begin{figure}[!htb]
    \centering
    \includegraphics[width=.85\textwidth]{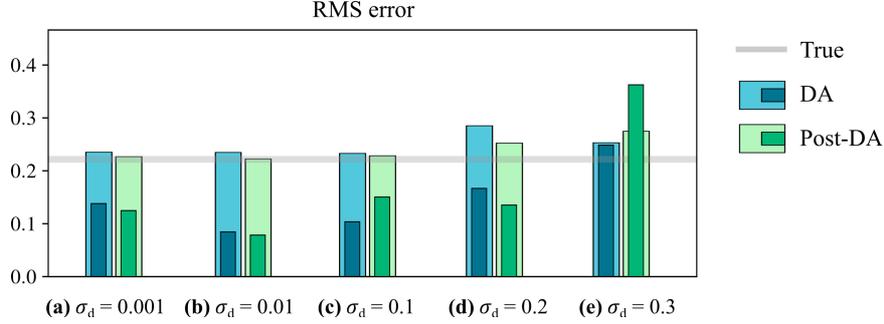}
    \caption{Nonlinear time-delayed model with a nonlinear bias.
    Analysis of the effect of increasing  the magnitude of the Gaussian additive noise in the assimilation. 
    Biased (thick bars) and unbiased (thin bars) RMS errors at the end of the assimilation (blues) and after the filter has been removed (greens).}
    \label{fig:gaussian_noise_levels}
\end{figure}

In the previous sections the observations were assumed to be subject to small-norm Gaussian noise with $\sigma_\mtxt{d}=0.01$ (see~\S~\ref{sec:Rijke}), because we assume that the sensors are well-calibrated.   
Figure~\ref{fig:gaussian_noise_levels} shows how relaxing this assumption affects the RMS errors at the end of the assimilation.  
We can see that the biased errors remain at true RMS levels up to a standard deviation in the observations of 10\% of the mean absolute amplitude. Moreover, the ESN successfully predicts the bias in the model with observation errors as high as 20\%.

\begin{figure}[!htb]
    \centering
    \includegraphics[width=\textwidth]{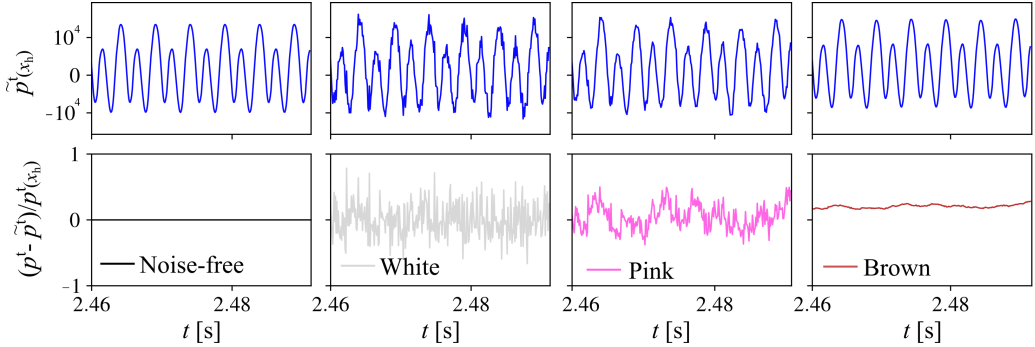}
    \caption{Nonlinear time-delayed model with a nonlinear bias.
    Comparison of the noise-free signal with the observation data with additive white, pink, and brown noises. Snapshots of the timeseries of the (a) acoustic pressure at the first microphone, with (b) their corresponding noises with $\mtxt{SNR_{dB}}=13.70\pm0.47$.}
    \label{fig:colors}
\end{figure}

\begin{figure}[!htb]
    \centering
    \includegraphics[width=\textwidth]{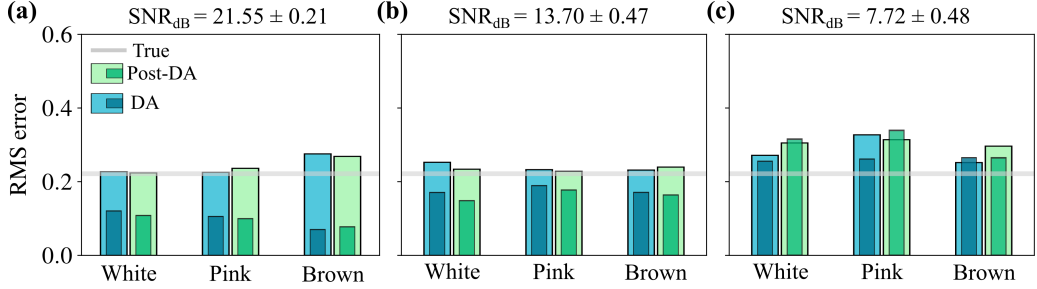}
    \caption{Nonlinear time-delayed model with a nonlinear bias.
    Performance of the r-EnKF with white, pink, and brown noises,  with (a) $\mtxt{SNR_{dB}}=21.55\pm0.21$, 
    (b) $\mtxt{SNR_{dB}}=13.70\pm0.47$, and 
    (c) $\mtxt{SNR_{dB}}=7.72\pm0.48$. 
    Biased (thick bars) and unbiased (thin bars) RMS errors at the end of the assimilation (blues) and after the filter has been removed (greens)
    }
    \label{fig:noise_colors_results}
\end{figure}

Finally, we analyse how the colour of the measurement noise affects the r-EnKF performance. We create white, pink and brown noises in frequency domain, normalize them to have equivalent energy spectrum, and then weight them with a factor of the mean absolute pressure. We consider three factors: 0.1, 0.25, and 0.5,  which correspond to a signal-to-noise ratio values of $\mtxt{SNR_{dB}} = 21.55\pm0.21, 13.70\pm0.47$, and $7.72\pm0.48$. We compute the signal-to-noise ratio as
\begin{equation}\label{eq:SNRdB}
    \mtxt{SNR_{dB}}=\dfrac{1}{N_q}\sum_{q=1}^{N_q}10\log_{10}{\dfrac{\left\langle {p^t_q}^2\right\rangle}{\left\langle {\left(p_q-\tilde{p}_q\right)}^2\right\rangle}}.
\end{equation}
Figure~\ref{fig:colors} shows snapshots of the three coloured noises for the factor 0.25, as well as the corresponding noise-free data. 
The results for the three noise levels are shown in Fig.~\ref{fig:noise_colors_results}. 
The r-EnKF successfully recovers solutions with small biased RMS up to SNR of approximately 14~dB, which is noteworthy because the assumption of Gaussian observation error remains in the assimilation, i.e.,~the filter has no information about the colour of the noise. These results are key for future application of the r-EnKF in real experimental data, because experimental devices are typically affected by non-Gaussian errors. 

\section{Conclusions}\label{sec:conclusions}
\toLM{
% Intro -----------------------------
Upon making physical assumptions and numerical approximations, high-fidelity physical models result in low-order models, which may be affected by uncertainties in the state, parameters, and by model biases. 
If the model is biased, data assimilation methods can provide state and parameters that are inaccurate, which make the model quantitatively incorrect and not generalizable. 
% 
% What the problem is -----------------------------
Current bias-aware real-time data assimilation methods face two limitations: 
(i) they cannot ensure that the model bias of the analysis is unique or small in norm;  and 
(ii)  they need an \textit{a priori} parametrized model for the bias, whose parameters are inferred within the assimilation. 
%
% What we do -----------------------------
To overcome these limitations, 
first, we propose a regularized bias-aware sequential assimilation framework, within we derive the regularized bias-aware ensemble Kalman filter (r-EnKF), which is designed to favour small-norm bias analyses, thus, bypassing (i); and 
second, we infer the model bias using echo state networks (ESNs), which overcome (ii) because they are universal approximators, and we design a multi-parameter training approach for the network.
}

\toLM{
% Results -----------------------------
%
We test and validate the proposed r-EnKF using two models of nonlinearly coupled oscillators (with and without time-delay), affected by different forms of bias (linear, nonlinear, and time-dependent functions of the state).  
The results show that the r-EnKF successfully infers a combination of model parameters that  provides a physical state with a unique bias, which can be accurately recovered by the ESN. 
Further,  we show that the designed multi-parameter training approach with data augmentation makes the ESN robust and flexible, because the same network can infer biases with different dynamics. 
We show that the weight of the bias regularization in the assimilation is key to recover small-norm bias analyses because it directly controls the magnitude of the instantaneous changes in state and parameters in every analysis step. 
Therefore, future work can be focused on the possibility of adaptively tuning the bias regularization weight factor.  
Finally, we show that the r-EnKF is robust to work with high noise observations as well as with measurements subject to colour noise.  
% Happy future ------------------------------
The proposed regularized bias-aware ensemble Kalman filter opens  new opportunities for real-time prediction in nonlinear systems by performing state, parameter, and bias estimation when data from sensors is available. 
}

\begin{description}
\item[\textbf{Code availability}] \url{https://github.com/MagriLab/rBA-EnKF}

\item[\textbf{Acknowledgements}]
A. N. is supported partly by the EPSRC-DTP, the Cambridge Commonwealth, European \& International Trust under a Cambridge European Scholarship, and Rolls Royce. A. R. is supported by the Eric and Wendy Schmidt AI in Science Postdoctoral Fellowship, a Schmidt Futures program.  L. M. gratefully acknowledges financial support from the ERC Starting Grant PhyCo 949388. L.M and A.N acknowledge support from the UKRI AI for Net Zero grant EP/Y005619/1. 
\end{description}

\newpage

%% The Appendices part is started with the command \appendix;
%% appendix sections are then done as normal sections
\appendix

\section{Nomenclature}\label{app:nomenclature}
\printnomenclature

\section{Recycle Validation}\label{app:ESN_RV}

Following \citet{racca_robust_2021}, we optimize the hyperparameters of the ESN \eqref{eq:ESN_OG} (the input scaling, $\sigma_\mtxt{in}$, and spectral radius, $\rho$) through Bayesian optimization and recycle validation (RV). 
In the RV, the optimal hyperparameters are chosen by minimizing the mean square error of closed-loop predictions over multiple intervals seen during training in open-loop. The number of intervals is known as the number of folds, $N_\mtxt{folds}$. Specifically, in each of the time series, we select  $N_\mtxt{folds}=4$ of length $t_\mtxt{val}$, which is problem-specific and reported in \ref{app:params}. 
The recycle validation algorithm exploits information from multiple parts of the dataset, whilst keeping a low computational cost \citep{racca_robust_2021}, because RV performs the validation on $N_\mtxt{folds}$ intervals used in training by keeping a fixed output matrix, $\matr{W}_\mtxt{out}$, but using the closed-loop configuration  rather than the open-loop setting used during training. 
The mean square error computed through RV is optimized in a hyperparameter space (whose ranges are reported in \ref{app:params}), through Bayesian optimization. 
We evaluate the error in 16 grid points of the hyperparameter space,  and then on 4 additional points, which are selected through the gp-hedge algorithm~\citep{brochu2010tutorial, hoffman2011portfolio}. 
% The procedure is implemented using NumPy \ar{??} and scikit-optimize.  

\section{Derivation of the Jacobian of the ESN}\label{app:ESN_Jac}
The Jacobian of ESN~\eqref{eq:ESN_OG} 
is the gradient of the output $\vect{b}^\mtxt{f}$  with respect to the input $\vect{i}$
\begin{equation}
    \dfrac{\totald\vect{b}^\mtxt{f}(t_{k+1})}{\totald\vect{i}(t_k)},
\end{equation}
where the input $\vect{i}(t_k)$ in open-loop is data, whereas in the closed-loop the ESN output at the previous time step $\vect{b}^\mtxt{f}(t_{k})$ is used as the input. 
We follow the procedure in~\S~\ref{sec:ESNJacobian} to rewrite~\eqref{eq:ESN_OG} as
\begin{align}\nonumber
        \vect{b}^\mtxt{f}(t_{k+1}) &= \matr{W}_\mtxt{out}^{(1)}\tanh\left(\sigma_\mtxt{in}\matr{W}_\mtxt{in}^{(1)}\left(\vect{i}(t_k)\odot\vect{g}\right)+\sigma_\mtxt{in}\delta_\mtxt{r}\vect{W}_\mtxt{in}^{(2)} +\rho\matr{W}\vect{r}(t_k)\right)+ \vect{W}_\mtxt{out}^{(2)}  \\
        \vect{r}(t_{k}) &= \left({\matr{W}_\mtxt{out}^{(1)}}\T \matr{W}_\mtxt{out}^{(1)} \right)^{-1}{\matr{W}_\mtxt{out}^{(1)}}\T\left(\vect{b}^\mtxt{f}(t_k)-\vect{W}_\mtxt{out}^{(2)}\right).
\end{align}
With this, the Jacobian becomes
\begin{equation}
    \dfrac{\totald\vect{b}(t_{k+1})}{\totald\vect{i}(t_k)} = \matr{W}_\mtxt{out}^{(1)}\left(\matr{T}\odot\left(\sigma_\mtxt{in}\matr{W}_\mtxt{in}^{(1)}\odot\matr{G}+\dfrac{\totald \vect{r}(t_k)}{\totald \vect{i}(t_k)}\right)\right), 
\end{equation}
where the matrices $\matr{G}$ and $\matr{T}$ are detailed in~\S~\ref{sec:ESNJacobian}.  
The reservoir state $\vect{r}(t_k)$ is a function of the output from the previous step,  $\vect{b}^\mtxt{f}(t_k)$.  In closed-loop, $\vect{i}(t_k)=\vect{b}^\mtxt{f}(t_k)$, and hence 
\begin{equation}
    \dfrac{\totald \vect{r}(t_k)}{\totald \vect{i}(t_k)} =  
    \begin{cases}
      \rho\matr{W}\left({\matr{W}_\mtxt{out}^{(1)}}\T \matr{W}_\mtxt{out}^{(1)} \right)^{-1}{\matr{W}_\mtxt{out}^{(1)}}\T & \text{if closed-loop}. \\
      \matr{0} & \text{if open-loop}.
    \end{cases}
\end{equation}
% \begin{subequations}
% \begin{empheq}[left={\dfrac{\totald\vect{b}(t_{k+1})}{\totald\vect{i}(t_k)} = \empheqlbrace\,}]{align}
% &\matr{W}_\mtxt{out}^{(1)}\left[\matr{T}\odot\left(\sigma_\mtxt{in}\matr{W}_\mtxt{in}^{(1)}\odot\matr{G}+\rho\matr{W}\left({\matr{W}_\mtxt{out}^{(1)}}\T \matr{W}_\mtxt{out}^{(1)} \right)^{-1}{\matr{W}_\mtxt{out}^{(1)}}\T \right)\right] &\text{if closed-loop}\label{eq:J_closed_app} \\[1em]
% &\matr{W}_\mtxt{out}^{(1)}\left[\matr{T}\odot\left(\sigma_\mtxt{in}\matr{W}_\mtxt{in}^{(1)}\odot\matr{G}\right)\right] &\text{if open-loop} \label{eq:J_open_app}
% \end{empheq}
% \end{subequations}
% 
% where $\matr{G}=\left[\vect{g}~|\dots|~\vect{g}\right]\T\in\mathbb{R}^{N_q\times N_r}$; 
% $\vect{1}\in\mathbb{R}^{N_r}$ and $\matr{0}\in\mathbb{R}^{N_r\times N_q}$ are a vector of ones and a matrix of zeros, respectively; 
% and  $\matr{T}=\left[\vect{T}~|\dots|~\vect{T}\right]\in\mathbb{R}^{N_r\times N_q}$ with
% \begin{align}\nonumber
% \vect{T}=\vect{1}-\tanh^2\left(\sigma_\mtxt{in}\matr{W}_\mtxt{in}^{(1)}\left(\vect{b}(t_k)\odot\vect{g}\right)+\sigma_\mtxt{in}\delta_\mtxt{r}\matr{W}_\mtxt{in}^{(2)} +\rho\matr{W}\vect{r}(t_k)\right),  
% \end{align}

% ---------------------------------------------------------------------------
\section{Nonlinear time-delayed model}\label{app:Rijke}
\begin{figure}[!htb]
    \centering
    \includegraphics[width=\textwidth]{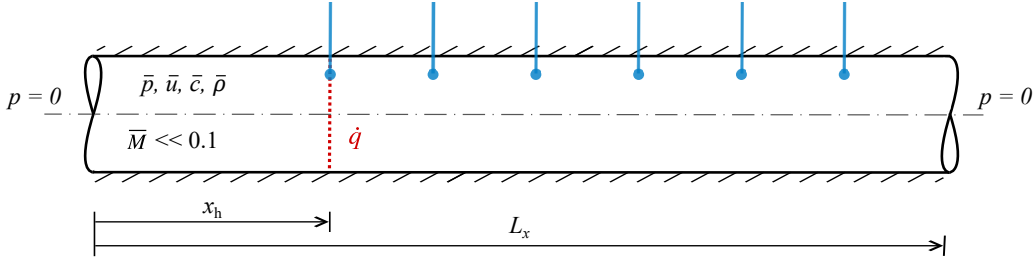}
    \caption{
    Schematic of the nonlinear time-delayed model, which consists of an open-ended tube with a compact heat source. 
    The measurable quantity  is the acoustic pressure at different locations in the tube, which are measured by microphones. 
    The blue vertical lines indicate the equidistant microphones, and the heat released by the compact heat source is indicated by the vertical dotted red line. }
    \label{fig:Rijke_drawing}
\end{figure}

The nonlinear time-delayed system analysed in~\S~\ref{sec:Rijke} is illustrated in Fig.~\ref{fig:Rijke_drawing}. The system models the thermoacoustic oscillations occurring in a laboratory-scale  open-ended tube  with a compact heat source (also known as the Rijke tube).  
We employ the low-order model detailed by~\citet{juniper2011}. 
The main modelling assumptions are: 
(i) the mean flow is a perfect gas with a small Mach number and constant properties, such that mean flow effects are neglected in the acoustic propagation; 
(ii) the boundary conditions are ideal, i.e. the acoustic pressure is zero at both ends of the tube, as illustrated in Fig.~\ref{fig:Rijke_drawing}; 
(iii) the viscous and body forces are negligible; and 
(iv) the heat source is compact and can be modelled as a point source of sound. 
Thus, the acoustics in the duct are governed by the one-dimensional linearized  momentum and energy conservation equations. 
The Markovian formulation introduced by~\citet{huhn_stability_2019} for the acoustic  velocity $u$ and pressure $p$ is 
\begin{subequations}\label{eq:sys_up}
\begin{align}
    &\dfrac{\partial u}{\partial t} + \dfrac{1}{\Bar{\rho}}\dfrac{\partial p}{\partial x} = 0 \\
    &\dfrac{\partial p}{\partial t} + \Bar{\rho}\,\Bar{c}^2\dfrac{\partial u}{\partial x} = \Dot{q}(\Gamma -1)\delta(x - x_\mtxt{h})  -\zeta \dfrac{\overbar{c}}{L_x}p\\
    \label{eq:advection}
    &\dfrac{\partial w}{\partial t} - \tau_\nu\dfrac{\partial w}{\partial X} = 0 %\quad,\qquad0\leq X\leq1
\end{align}
\end{subequations}
where 
$x_\mtxt{h}$ is the heat source location;  
$\overbar{\rho}, \overbar{c}$ and $\overbar{u}$ are the mean-flow density, speed of sound and velocity, respectively; 
$\Gamma$ is the heat capacity ratio; 
$\zeta$ is the damping factor; 
${\delta}$ is the Dirac delta distribution, which enforces a compact assumption for the heat release rate $\Dot{q}$ [W/m$^2$],  which we model with the simple time-delayed square-root law
\begin{equation}\label{eq:qdot_app}
    \Dot{q}(t)= \overbar{u}\overbar{p}\beta\left(\sqrt{\left|{\dfrac{1}{3}+\dfrac{u\left(x_\mtxt{h}, t-\tau\right)}{\overbar{u}}}\right|}-\sqrt{\dfrac{{1}}{3}}\right), 
\end{equation}

The advection equation \eqref{eq:advection} transforms the time-delayed problem into an initial value problem. The equation is used to keep track of the acoustic velocity history at the heat source location, i.e. it acts as a numerical memory. 
The dummy variable $w$, which  travels in a dimensionless spatial domain $X$ with velocity $-\tau_\nu$, models the history of the acoustic velocity at the heat source location, $x_\mtxt{h}$. 
 We define the boundary condition
$
    w\left(X = 0,\,t\right) = u\left(x=x_\mtxt{h}, t\right),
$ 
such that the acoustic velocity at the heat source location at some time $t-\Delta$ is given by
$
    u\left(x_\mtxt{h},t-\Delta\right)= w\left(0,\,t-\Delta \right)=w\left(\tau_\nu/\Delta,\,t\right). 
$
The velocity of the advection is $\tau_\nu\geq\tau$, because the length of the stored history must be at least equal to the acoustic time delay. (Note that if $\tau$ is constant, one can set $\tau_\nu=\tau$,   
such that $u\left(x_\mtxt{h}, t-\tau\right)=w\left(1,t\right)$.)

The acoustic equations are discretized in space with a Galerkin method by taking the natural acoustic modes as the orthogonal spatial basis. 
The ideal boundary conditions ($p(x=0) = p(x = L_x) =0$) are enforced by choosing for the velocity and pressure oscillations the basis
\begin{align}
    u(x,t)&=\sum^{N_m}_{j=1}\,\eta_j(t)\,\cos{\left(\dfrac{\omega_j}{\Bar{c}} x\right)}\quad \mtxt{and} \quad
    p(x,t)=-\sum^{N_m}_{j=1}\,\mu_j(t)\sin{\left(\dfrac{\omega_j}{\Bar{c}} x\right)},
\end{align}
where $N_m$ is the number of 
 the acoustic modes. 
Finally, we discretize the advection equation in $X$ with a Chebyshev spectral method, using the  Chebyshev polynomial matrix $\matr{D}$ for $N_c+1$ collocation points, which are defined as $X_i=0.5\left(1-\cos{\left({i\pi/N_c}\right)}\right)$ for $i=0,\dots,N_c$. 
Introducing the Galerkin and Chebyshev discretizations into \eqref{eq:sys_up} yields the governing equations
\begin{subequations}\label{eq:sys_modes_app}
\begin{align}
    &\dfrac{\mtxt{d}\eta_j}{\mtxt{d} t} = \dfrac{\omega_j}{\Bar{\rho}\Bar{c}}\mu_j \\
    &\dfrac{\mtxt{d} \mu_j}{\mtxt{d} t} = - \Bar{\rho}\,\Bar{c}\,\omega_j\eta_j -2\,\Dot{q}\dfrac{\Gamma-1}{L_x}\sin{\left(\dfrac{\omega_j}{\Bar{c}}x_\mtxt{h}\right)}- \zeta_j \dfrac{\bar{c}}{L_x} \mu_j, \quad  \mtxt{for} \quad j=0,\dots,N_m-1
    \\
    &\dfrac{\mtxt{d}\nu_i}{\mtxt{d}t} = 2\tau_\nu \sum_k\tensor{D}_{ik} \nu_k \quad  \mtxt{for} \quad i=0,\dots,N_c-1
\end{align}
\end{subequations}
where 
$\eta_j$ and $\mu_j$ are the acoustic velocity and pressure modes resulting from the Galerkin discretization of the acoustic velocity, $u$, and pressure, $p$, into $N_m$ acoustic modes with $\omega_j$ acoustic frequencies; 
$\zeta_j$ is the damping, defined with the modal form $\zeta_j=C_1 j^2 + C_2 \sqrt{j}$~\citep{LANDAU_1987}.

\section{Simulations' parameters}\label{app:params}

This appendix summarize the parameters used for the van der Pol (Tab.~\ref{tab:VdP_params}) and nonlinear time-delayed model (Tab.~\ref{tab:Rijke_params}) simulations.
\begin{table}[!hbt]
\centering
\caption{Parameters used in the van der Pol simulations. The symbol ($\dag$) indicates that it is the default value but it may be varied in different simulations; the asterisk ($^*$) indicates that the parameter is optimized within the given values; and if no units are specified, the quantity is dimensionless.}
\label{tab:VdP_params}
\begin{tabular}{|c|lr@{~}l|lr@{~}l|lr@{~}l|}
\hline
{\parbox[t]{2mm}{\multirow{4}{*}{\rotatebox[origin=c]{90}{Model}}}} & $\Delta t$ & 1\E{-4} & s & $\omega$ & $240\pi$ & Hz & $\beta_\mtxt{limits}$ & (20, 120) & Hz \\
 & $\overbar{\beta}^0$ & 70.0 & Hz & $\beta^\mtxt{t}$ & 75.0 & Hz &  $\kappa_\mtxt{limits}$ & (0.1, 10.0) &\\
 & $\overbar{\kappa}^0$ & 4.0 &  & $\kappa^\mtxt{t}$ & 3.4 &  &  $\zeta_\mtxt{limits}$ & (20, 120) & Hz \\
 & $\overbar{\zeta}^0$ & 60.0 & Hz & $\zeta^\mtxt{t}$ & 55.0 & Hz &  & & \\\hline
{\parbox[t]{2mm}{\multirow{3}{*}{\rotatebox[origin=c]{90}{Filter}}}} & $\Delta t_\mtxt{d}$ & $30\Delta t$ & s & $N_q$ & 1 &  & $\sigma_\mtxt{d}$ & 0.01 &  \\
 & $m$ & 10 &  & $t_\mtxt{err}$ & 0.04 & s & $\sigma_\alpha$ & 0.25 &   \\
  &  &  &  &  &  & &  $\sigma_\phi$ & 0.25 &  \\\hline
{{\parbox[t]{2mm}{\multirow{4}{*}{\rotatebox[origin=c]{90}{ESN}}}}} & $\Delta t_\mtxt{ESN}$ & $5 \Delta t$ & s & $L^\dag$ & 10 &  & $\lambda$ & $10^{-16}$ &  \\
 & $N_r$ & 100 &  & $\sigma_L$ & 0.5 &  & $\sigma_\mtxt{in}^*$ & $[10^{-5},1]$ &  \\
 & $N_\mtxt{wash}$ & 30 &  & $t_\mtxt{tr}$ & 1.0 & s & $\rho^*$ & $[0.7, 1.05]$ &  \\
 & $N_\mtxt{folds}$ & 4 &  & $t_\mtxt{val}$ & 0.01 & s &  &  & \\
\hline
\end{tabular}
\end{table}
\begin{table}[!hbt]
\centering
\caption{Parameters used in the nonlinear time-delayed model simulations. The symbol ($\dag$) indicates that it is the default value but it may be varied in different simulations; the asterisk ($^*$) indicates that the parameter is optimized within the given values; and if no units are specified, the quantity is dimensionless.}
\label{tab:Rijke_params}
\begin{tabular}{|c|lr@{~}l|lr@{~}l|lr@{~}l|}
\hline
{\parbox[t]{2mm}{\multirow{6}{*}{\rotatebox[origin=c]{90}{Model}}}} & $\Delta t$ & 1\E{-4} & s & $C_1$ & 0.05 &  & $\overbar{\beta}^0$ & 4.0 &  \\
 & $N_c$ & 50 &  & $C_2$ & 0.01 &  & $\overbar{\tau}^0$ & 1.5\E{-3} & s \\
 & $N_m$ & 10 &  & $\overbar{u}$ & 10 & m/s & $\beta^\mtxt{t}$ & 4.2 &  \\
 & $\tau_\nu$ & $10^{-2}$ & s & $\overbar{p}$ & 1.013 & bar & $\tau^\mtxt{t}$ & 1.4\E{-3} & s \\
 & $x_\mtxt{h}$ & 0.2 & m & $\overbar{T}$ & 417.2 & K & $\beta_\mtxt{limits}$ & (0.1, 5.0) &  \\
 & $L_x$ & 1.0 & m & $\Gamma$ & 1.4 &  & $\tau_\mtxt{limits}$ & (1\E{-6}, $\tau_\nu$) & s \\\hline
\parbox[t]{2mm}{\multirow{4}{*}{\rotatebox[origin=c]{90}{Filter}}} & $\Delta t_\mtxt{d}^\dag$ & $20\Delta t$ & s & $N_q$ & 6 &  & $\sigma_\mtxt{d}$ & 0.01 &  \\
 & $m^\dag$ & 50 &  & $\vect{x}$ & [0.2, 0.33, 0.47, &  & $\sigma_\alpha$ & 0.2 &  \\
 & $t_\mtxt{err}$ & 0.02 & s &  & 0.6, 0.73, 0.87] & m & $\sigma_\phi$ & 0.2 &  \\\hline
{\parbox[t]{2mm}{\multirow{3}{*}{\rotatebox[origin=c]{90}{ESN}}}} & $\Delta t_\mtxt{ESN}$ & $2 \Delta t$ & s & $L^\dag$ & 50 &  & $\lambda$ & $10^{-16}$ &  \\
 & $N_r$ & 500 &  & $\sigma_L$ & 0.2 &  & $\sigma_\mtxt{in}^*$ & $[10^{-5},10^{-2}]$ &  \\
 & $N_\mtxt{wash}$ & 50 &  & $t_\mtxt{tr}$ & 0.5 & s & $\rho^*$ & $[0.7, 1.05]$ &  \\
 & $N_\mtxt{folds}$ & 4 &  & $t_\mtxt{val}$ & 0.02 & s &  &  &  \\ \hline
\end{tabular}
\end{table}

% --------------------------------------------------------------------
\newpage
\section{Parameter convergence results}\label{app:params_convergence}
\toLM{
Figure~\ref{fig:errors_VdP_params} shows the parameter distribution at the end of the assimilation for simulations in Fig.~\ref{fig:errors_VdP}. 
The filter provides solutions with large uncertainties for  $L=1,~\mtxt{and}~10$ (as seen by the large ensemble spread in Figs.~\ref{fig:errors_VdP_params}{a,b}), which indicate non-converged solutions.
Further, we see that there are many ($\beta, \zeta, \kappa$) combinations that recover `good' unbiased analysis, which is expected as there are multiple solutions in the van der Pol system (see Eq.~\eqref{eq:VdPwave}). 
\begin{figure}[!htb]
    \centering
    \includegraphics[width=.95\textwidth]{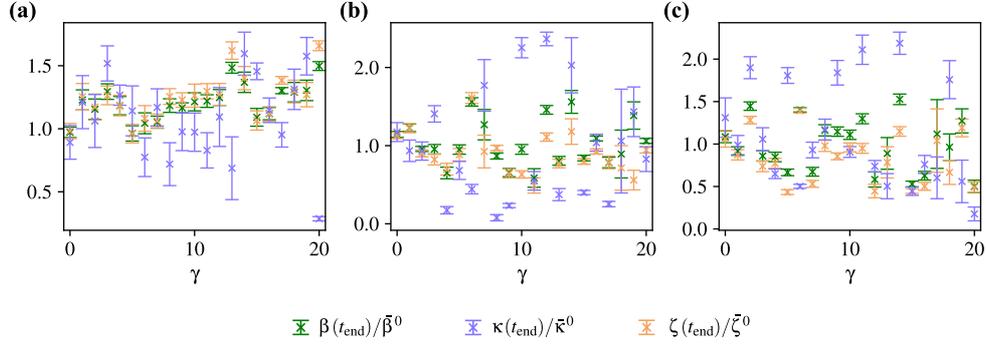}
    \caption{Van der Pol model. 
    Distribution of the inferred model parameters with their uncertainties at the end of the assimilation. Results for varying bias regularization factor $\gamma$, for different numbers of training sets 
    (a) $L=1$, (b) $L=10$, and (b) $L=50$. 
    The parameters are normalized by the initial ensemble mean.}
    \label{fig:errors_VdP_params}
\end{figure}
}

\toLM{
Figure~\ref{fig:Rijke_P} shows the parameter distribution at the end of the assimilation in Fig.~\ref{fig:Rijke_CRP}. We can see that, in contrast to Fig.~\ref{fig:errors_VdP_params}, the filter consistently converges to a similar set of parameters when the biased and unbiased errors are small, i.e.,~at the range of bias regularization factor $1\lesssim \gamma \lesssim 5$  (see Fig.~\ref{fig:Rijke_CRP}).
% In addition, Fig.~\ref{fig:Rijke_CRP}{a} shows that the filter converges to the same parameters when the minimum bias solution is recovered  (the small RMS values in Fig.~\ref{fig:Rijke_CRP}{a} at $\gamma\in[1.5, 4.5]$  correspond to similar $\vect{\alpha}$ combinations in Fig.~\ref{fig:Rijke_CRP}{a}).
\begin{figure}[!htb]
    \centering
    \includegraphics[width=.9\textwidth]{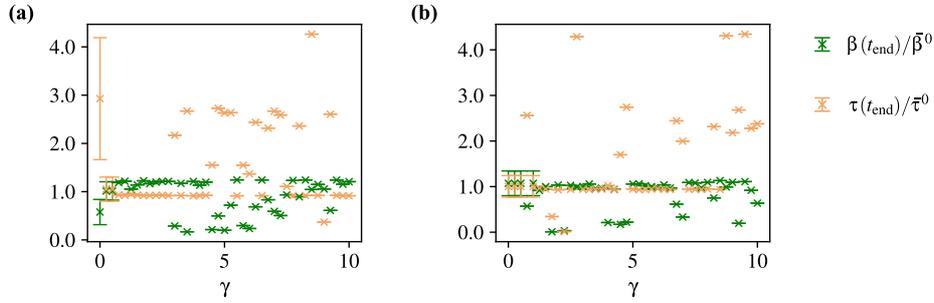}
    \caption{Nonlinear time-delayed model. 
    Distribution of the inferred model parameters with their uncertainties at the end of the assimilation. Results for varying bias regularization factor $\gamma$, for the (a) linear and (b) nonlinear bias cases at fixed $L=70$ and $m=50$. 
    The parameters are normalized by the initial ensemble mean.}
    \label{fig:Rijke_P}
\end{figure}
}
% --------------------------------------------------------------------

\section{Analysis of  the ensemble size}\label{app:m10-80}

\begin{figure}[!htb]
    \centering
    \includegraphics[width=\textwidth]{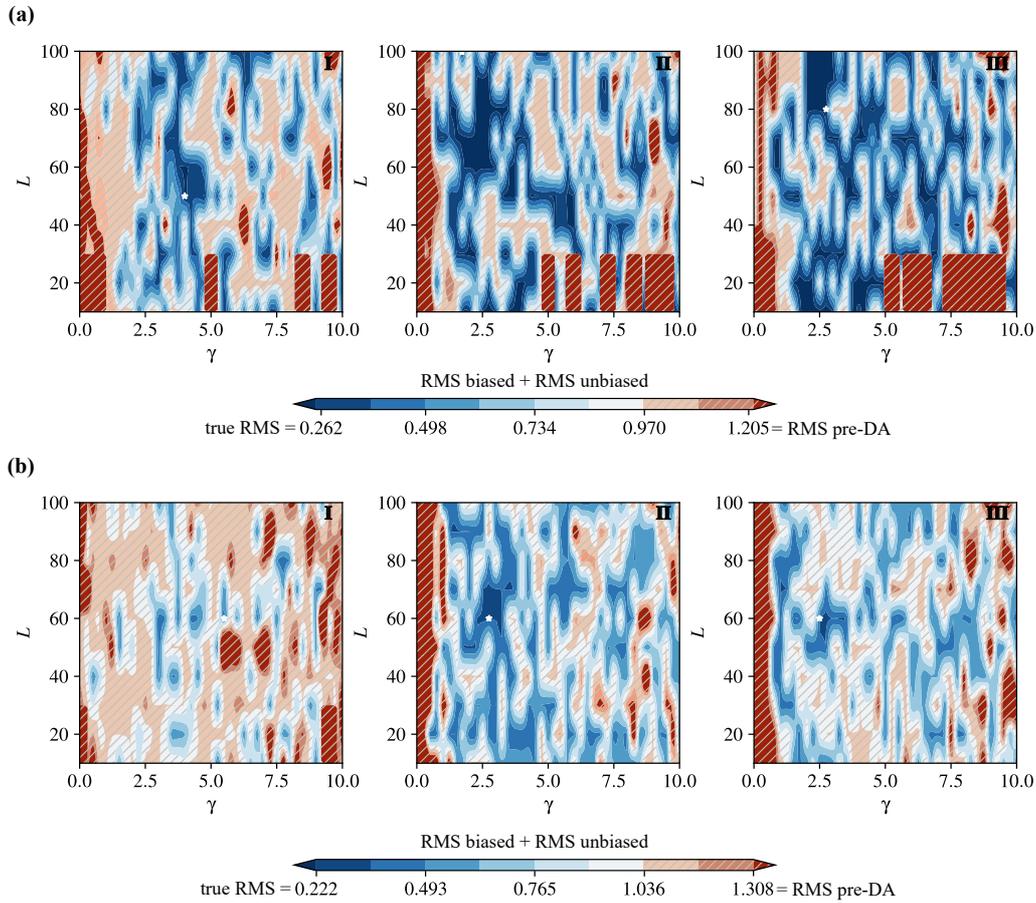}
    \caption{Nonlinear time-delayed model.  Error comparison of the post-assimilation error metrics for three ensemble sizes  (I) $m=10$, (II) $m=50$, (III) $m=80$, for the (a) linear and (b) nonlinear biases. 
    Contours for varying training set size, $L$, and bias regularization factor, $\gamma$. 
    The colourmap shows the sum of the bias and unbiased RMS errors, and the minimum combined RMS point is indicated with a white star.  } 
    \label{fig:Contours_m10-80}
\end{figure}

Figure~\ref{fig:Contours_m10-80} shows the error contours for three ensemble sizes for the nonlinear time-delayed model with linear and nonlinear biases.
The difference between $m=10$ and $50$ is significant, whilst increasing to $80$ does not notably vary the results. Thus, we fix $m=50$ in~\S~\ref{sec:Rijke_results}. 
% 
% --------------------------------------------------------------------

\nomenclature[A,a$\alpha$]{$\vect{\alpha}$}{Model parameters}
\nomenclature[A,b]{$\vect{b}$}{Model bias}
\nomenclature[A,Cbb]{$\matr{C}_{bb}$}{Bias error covariance matrix}
\nomenclature[A,Cdd]{$\matr{C}_{dd}$}{Observation error covariance matrix}
\nomenclature[A,Cpp]{$\matr{C}_{\psi\psi}$}{Ensemble error covariance matrix}
\nomenclature[A,d]{$\vect{d}$}{Observable data}
\nomenclature[A,Dtd]{$\Delta t_\mtxt{d}$}{Time between observables}
\nomenclature[A,Dt]{$\Delta t$}{Time step}

\nomenclature[A,e$\epsilon$]{$\vect{\epsilon}$}{Gaussian noise}
\nomenclature[A,F]{$\mathcal{F}$}{Nonlinear differential operator}
\nomenclature[A,g$\gamma$]{$\gamma$}{Bias regularization parameter}

\nomenclature[A,Id]{$\mathbb{I}_n$}{Identity matrix of $n\times n$ dimensions}
\nomenclature[A,J]{$\matr{J}$}{Jacobian (gradient of the bias with respect to the analysis state)}
\nomenclature[A,J]{$\mathcal{J}$}{DA cost function}

\nomenclature[A,M$\mathcal{M}$]{$\mathcal{M}$}{Nonlinear measurement operator}
\nomenclature[A,M]{$\matr{M}$}{Linear measurement operator}
\nomenclature[A,Mpsi]{$\matr{M}\vect{\psi}$}{Biased model prediction}
\nomenclature[A,m]{$m$}{Ensemble size}

\nomenclature[A,N$\matcal{M}$]{$\mathcal{N}$}{Normal random distribution}
\nomenclature[A,N]{$N$}{Augmented state vector length}
\nomenclature[A,Na]{$N_\alpha$}{Number of inferred parameters}
\nomenclature[A,Nphi]{$N_\phi$}{Number of state variables}
\nomenclature[A,Nq]{$N_q$}{Number of observables}

\nomenclature[A,p$\phi$]{$\vect{\phi}$}{State variables}
\nomenclature[A,p$\psi$]{$\vect{\psi}$}{Augmented state vector}

\nomenclature[A,s$\sigma_\alpha$]{$\sigma_\alpha$}{Ensemble parameters' initial standard deviation}
\nomenclature[A,s$\sigma_\phi$]{$\sigma_\phi$}{Ensemble states' initial standard deviation}

\nomenclature[A,t$t_err$]{$t_\mtxt{err}$}{Error metrics time window}

\nomenclature[A,U]{$\mathcal{U}$}{Uniform random distribution}
\nomenclature[A,y]{$\vect{y}$}{Unbiased model prediction}

% ESN

% \nomenclature[B,btr]{$\vect{b}^\mtxt{tr}$}{Training bias data}
% \nomenclature[B,btrn]{$\tilde{\vect{b}}^\mtxt{tr}$}{Noisy training data}
\nomenclature[B,D$\Delta$]{$\Delta t_\mtxt{ESN}$}{ESN time step}
\nomenclature[B,d$\delta$]{$\delta_\mtxt{r}$}{Symmetry-breaking constant}

\nomenclature[B,g]{$\vect{g}$}{Input normalizing factors}

\nomenclature[B,L]{$L$}{Number of training datasets}
\nomenclature[B,l$\lambda$]{$\lambda$}{Tikhonov parameter}
\nomenclature[B,Ntr]{$N_\mtxt{tr}$}{Length of the training set}
\nomenclature[B,Nr]{$N_r$}{Number reservoir states (neurones)}

\nomenclature[B,r]{$\vect{r}$}{Reservoir state vector}
\nomenclature[B,r$\rho$]{$\rho$}{Spectral radius}

\nomenclature[B,s$\sigma$]{$\sigma_\mtxt{in}$}{Input scaling}

\nomenclature[B,ttr]{$t_\mtxt{tr}$}{Training time}
\nomenclature[B,tval]{$t_\mtxt{val}$}{Validation time}
\nomenclature[B,W]{$\matr{W}$}{Reservoir state matrix}
\nomenclature[B,Win]{$\matr{W}_\mtxt{in}$}{Input matrix}
\nomenclature[B,Wout]{$\matr{W}_\mtxt{out}$}{Output matrix}

% \nomenclature[A,zj]{$()_j$}{Ensemble member index}
% \nomenclature[A,zq]{$()_q$}{Observable dimension index}
% \nomenclature[A,zk]{$()_k$}{Time step index}
% \nomenclature[A,zaa]{$()^\mtxt{a}$}{Analysis}
% \nomenclature[A,zat]{$()^\mtxt{t}$}{True}
% \nomenclature[A,zaf]{$()^\mtxt{f}$}{Forecast}
% \nomenclature[B,ztr]{$()^\mtxt{tr}$}{Train}

% \newpage
\bibliographystyle{plainnat} 
\bibliography{main}

\end{document}